\documentclass[nonacm]{acmart}

\makeatletter
\newcommand{\myconfshort}{\acmConference@shortname}
\newcommand{\myconffull}{\acmConference@name}
\newcommand{\myconfdate}{\acmConference@date}
\newcommand{\myconfloc}{\acmConference@venue}
\AtBeginDocument{
  \fancypagestyle{firstpagestyle}{
    \fancyhead{}%
    \fancyfoot[C]{}%
  }
  \fancyhf{}
  \fancyhead[LO]{\@headfootfont\shorttitle}%
  \fancyhead[RE]{\@headfootfont\@shortauthors}%
  \fancyhead[LE]{\@headfootfont\footnotesize \myconfshort, \myconfdate, \myconfloc}%
  \fancyhead[RO]{\@headfootfont\footnotesize \myconfshort, \myconfdate, \myconfloc}%
  \fancyfoot[C]{}%
}
\makeatother
\acmBooktitle{\conffull\@ (\confshort), \confdate, \confloc}

\usepackage{etoolbox}
\usepackage{color}
\usepackage[normalem]{ulem}
\usepackage{amsmath}
\newtoggle{comments}
\toggletrue{comments}
\usepackage{booktabs}
\usepackage{array}
\usepackage{adjustbox}
\usepackage{graphicx}
\usepackage{subcaption}
\usepackage[justification=centering]{caption}
\usepackage{comment}
\usepackage[flushleft]{threeparttable}
\begin{document}

\title[Improving Accuracy and Fairness in Property Tax Assessments]{Tradeoffs are Domain Dependent: Improving Accuracy and Fairness in Property Tax Assessments}

\author{Evelyn Smith}
\authornote{Both authors contributed equally to this research.}
\email{esmith@abfn.org}
\orcid{0000-0003-2470-1533}
\affiliation{%
  \institution{American Bar Foundation}
  \city{Chicago}
  \state{Illinois}
  \country{USA}
}

\author{Emma Harvey}
\authornotemark[1]
\email{evh29@cornell.edu}
\orcid{0000-0001-8453-4963}
\affiliation{%
    \institution{Cornell Tech}
    \city{New York}
    \state{New York}
    \country{USA}
}

\author{Christopher Berry}
\email{crberry@uchicago.edu}
\orcid{0000-0001-6978-9183}
\affiliation{%
    \institution{University of Chicago}
    \city{Chicago}
    \state{Illinois}
    \country{USA}
}

\author{Jacob Goldin}
\email{jsgoldin@uchicago.edu}
\authornote{Equal co-supervising authors.}
\orcid{0000-0001-5518-0027}
\affiliation{%
    \institution{University of Chicago}
    \city{Chicago}
    \state{Illinois}
    \country{USA}
}
\author{Daniel E. Ho}
\email{deho@law.stanford.edu}
\authornotemark[2]
\orcid{0000-0002-2195-5469}
\affiliation{%
    \institution{Stanford University}
    \city{Stanford}
    \state{California}
    \country{USA}
}

\renewcommand{\shortauthors}{Smith and Harvey et al.}

\copyrightyear{2026}
\acmYear{2026}
\setcopyright{cc}
\setcctype{by}
\acmConference[FAccT '26]{The 2026 ACM Conference on Fairness, Accountability, and Transparency}{June 25--28, 2026}{Montreal, QC, Canada}
\acmBooktitle{The 2026 ACM Conference on Fairness, Accountability, and Transparency (FAccT '26), June 25--28, 2026, Montreal, QC, Canada}
\acmDOI{10.1145/3805689.3812318}
\acmISBN{979-8-4007-2596-8/2026/06}

\begin{abstract}
Algorithmic fairness research often assumes a tradeoff between fairness and accuracy. Yet this tradeoff may not be universal. We test this assumption in the context of U.S. property tax assessment\textemdash a setting in which the output of predictive algorithms directly determines the distribution of tax obligations among homeowners. Currently, systematic assessment errors cause owners of lower-valued properties to face disproportionately high tax burdens, creating regressivity in the property tax system. Using data on 26 million property sales spanning 95\% of U.S. counties, we conduct three complementary analyses. First, we find that assessment accuracy and fairness—measured using domain-relevant metrics—are strongly correlated across counties under status quo practices. Second, in simulated assessment models, we show that adding property features improves accuracy in most cases, and that when accuracy improves, fairness almost always improves as well. Third, we show that incorporating publicly available Census data into assessment models—a feasible reform in most counties—would significantly improve both accuracy and fairness relative to status quo assessments. Together, these results challenge the presumed universality of the fairness–accuracy tradeoff and demonstrate that well-designed modeling improvements can advance both fairness and accuracy in large-scale public sector systems.\looseness=-1
\end{abstract}




\maketitle

\section{Introduction}\label{s-1-introduction}

\enlargethispage{20pt}
For the past decade, algorithmic fairness researchers have grappled with the \textit{fairness-accuracy tradeoff} \cite{kleinberg_inherent_2017, corbett-davies_algorithmic_2017, pleiss_fairness_2017, menon_cost_2018}, empirically identified across domains including criminal justice \cite{chouldechova_fair_2016, compas}, child welfare \cite{chouldechova_case_2018}, hiring \cite{quinonero_candela_disentangling_2023}, and advertising \cite{koenecke_popular_2023}. The widespread presence of this tradeoff has led to an implicit assumption that it is inherent: that, in most real-world settings, building fairer models means forfeiting accuracy \cite{cooper_emergent_2021}.

Here, we present evidence from one domain\textemdash previously unstudied in the algorithmic fairness literature\textemdash where improvements in accuracy are accompanied by systematic and substantial gains in fairness: property taxes. The vast majority of Americans pay property taxes, which are an essential source of revenue for local governments and are used to pay for public services and infrastructure like schools, roads, and fire departments~\cite{proptaxlocal}. To collect a property tax, taxing jurisdictions (e.g., U.S. counties) \textit{assess} properties by determining their fair market value, and levy taxes in proportion to these assessed values less any caps, credits, or exemptions.\footnote{Assessments are distinct from \textit{appraisals}, which are estimates of the fair market value of a property for non-tax purposes, such as real estate transactions. However, both assessments and appraisals are estimates of market value\textemdash their key distinction lies in their application rather than their objective. If a property is over-assessed, the resulting tax burden will exceed what policymakers intended even after exemptions are applied. For this reason and consistent with prior literature, we evaluate accuracy and fairness relative to \textit{market value}, defined as the price a property actually sells for, and we discuss the limitations of this choice in S5.1} \looseness=-1

While the property tax is intended to be a \textit{proportional tax}~\cite{proptaxhistory}, multiple studies have shown that it is \textit{regressive} in practice: individuals who own lower-valued properties pay a higher proportion of their values in taxes than individuals who own higher-valued properties \cite{mcmillenAssessmentRegressivityProperty2020, berryReassessingPropertyTax2021, amornsiripanitchWhyAreResidential2022, avenancio-leonAssessmentGapRacial2022}. Regressivity disproportionately penalizes non-white Americans, who are more likely to own lower-valued properties due to both wealth inequality and historical segregation in the housing market \cite{abc_regressivity, harris2004assessing}. Because regressive property taxes are not proportional to home value, they violate the important tax policy principle of \textit{vertical equity}, which  requires that ``different individuals be treated appropriately differently'' \cite{black_algorithmic_2022}.\looseness=-1

Regressivity has many potential causes \cite{berryReassessingPropertyTax2021}. In some cases, it is driven by corruption or  malfeasance. For example, assessors in Cook County, Illinois accepted bribes in exchange for reducing assessments of high-value properties, prompting an FBI investigation and jail time for some conspirators \cite{ct_regressivity, ct_regressivity2, ct_regressivity3, ct_regressivity4, cst_fbi}. In other cases, legal restrictions on assessments drive regressivity. For example, 17 U.S. states place caps on assessment increases \cite{dornfestStateProvincialProperty, dornfest2014state, dornfest2019state, beebe2025texas}. These caps are intended to protect individual homeowners from large hikes in tax bills, but also shift tax obligations from homeowners whose properties have appreciated quickly to those whose values have grown at slower rates \cite{nyc_report, hayashi2014property}. Regressivity has even been leveraged as a tool to suppress social movements. For example, following boycotts and pickets organized by Black residents of Edwards, Mississippi in 1966, the local assessor increased the assessed values of Black-owned homes by ten times as much as white-owned homes, with the largest increases assigned to boycott participants and organizers. \cite{kahrl2024black}.\looseness=-1 

However, the most significant sources of assessment regressivity today are shortcomings in the data and models used to assess properties \cite{berryReassessingPropertyTax2021, amornsiripanitchWhyAreResidential2022}. Most assessors conduct simplistic assessments based on data that can be collected at scale \cite{bidansetSurveyUseAutomated2022}. In contrast, the fair market value of a property is defined according to its sale price in an arms-length transaction.\footnote{An \textit{arms-length transaction} is one where the buyer and seller are not known to one another.} In such transactions, buyers and sellers have access to a rich set of information, including in-person tours, inspections, and appraisals, which they use to negotiate a sale price. Because most assessors rely on comparatively sparse data, their assessments can regress to the mean market values in their jurisdiction. This results in low-value properties being, on average, over-assessed and over-taxed and high-value properties being, on average, under-assessed and under-taxed. Thus, regressivity occurs even in the absence of corruption or legal restrictions.\looseness=-1

In this paper, we assess the hypothesis that more accurate assessments generally lead to fairer (i.e., less regressive) property taxes\textemdash in other words, that property tax assessment is not typically subject to the fairness-accuracy tradeoff.
\enlargethispage{10pt}
To examine this hypothesis, we take a holistic view of real-world assessment pipelines and consider whether efforts to increase assessment accuracy generally lead to increases in fairness. In doing so, we make the following key contributions:\looseness=-1

\begin{itemize}
    \item We describe the relationship between accuracy and regressivity in the context of property tax assessments using real-world data on 26 million property sales and assessments spanning 95\% of U.S. counties over five years, finding that, across jurisdictions, assessment accuracy is positively associated with taxation fairness. This pattern suggests that methodological improvements in assessment practice—of which machine-learned approaches are a prominent example—may yield joint gains in accuracy and fairness.
    \item Using simulated assessment models, we find that adding additional property characteristics increases assessment accuracy in most cases, and that accuracy and fairness gains coincide more than 99\% of the time.
    \item We propose simple and broadly applicable computational approaches to improve assessment accuracy and fairness. In particular, we estimate that the addition of publicly available Census characteristics can significantly improve the accuracy and fairness of existing assessments in hundreds of U.S. counties, and can also reduce regressivity with respect to neighborhood-level race and income.
\end{itemize}

Our analysis builds on an established literature studying the distribution of assessment errors with respect to property value \citep[e.g.,][]{berryReassessingPropertyTax2021, amornsiripanitchWhyAreResidential2022, hodge2017assessment, lin2010property}, income \citep{mcmillen2020assessment, edelstein1979appraisal}, and race \citep{avenancio-leonAssessmentGapRacial2022}. Prior work has mostly explored the effects of specific changes in assessment method on both accuracy and regressivity within a single jurisdiction \citep[e.g.,][]{lin2010property, geraci1977measuring, hou2025assessment}.\footnote{An important exception is an independent working paper by \citet{amornsiripanitchWhyAreResidential2022}, which studies how incorporating the housing price index constructed by \citet{bayer2017racial} into property tax assessments affects regressivity and accuracy within a pooled sample of all U.S. taxing jurisdictions.} We contribute to this literature by studying the relevance of fairness-accuracy tradeoffs in practice, using data drawn from the universe of U.S. taxing jurisdictions and variation in algorithmic methods that are realistic and feasible for jurisdictions to implement. We also identify the conditions under which gains in assessment accuracy are likely to come at the expense of regressivity, and examine heterogeneity across counties in accuracy and fairness responses to changes in assessment methodology.\looseness=-1

We proceed as follows. In \S\ref{s-2-background}, we provide background on property tax assessments in the U.S. as well as on the fairness-accuracy tradeoff. We outline our methods in \S\ref{s-3-methods} and describe our empirical results in \S\ref{s-4-results}. We close by discussing the implications of our findings, avenues for future work, and limitations of our approach in \S\ref{s-5-discussion}.\looseness=-1
\section{Background and Related Work}\label{s-2-background}
\subsection{The United States Property Tax Assessment System}
In the U.S., property tax assessments are conducted at the local (e.g., county) level. Local assessors typically rely on one of three approaches to estimate the fair market value of a property. The \textit{cost} approach values properties according to how much it would cost to reproduce them, less depreciation; the \textit{sales} approach values properties  in line with similar, recently sold properties; and the \textit{income} approach values commercial or rental properties according to the income they produce~\cite{kokinis2005use, IAAO2018AVM}. Local governments have discretion to conduct assessments as they see fit, subject to state-level regulation of assessment processes or outcomes. For example, as of 2025, 17 states required that local assessors cap year-over-year assessment increases; 27 states required that assessors conduct in-person property inspections as part of their assessments; and 39 states imposed limits on the time between reassessments \citep{beebe2025texas, dornfest2019state, dornfest2014state}.\looseness=-1

While researchers have repeatedly found that regression-based assessments, which are also referred to as Automated Valuation Models (AVMs), outperform alternative methods of assessment ~\cite{jennifer2021garbage, bidanset2017accounting, quintos2014improving, droj2024comprehensive}, a 2019 International Association of Assessing Officers (IAAO) survey found that just 16\% of assessor's offices use AVMs \cite{bidansetSurveyUseAutomated2022}.\footnote{The vast majority of assessor's offices do not release information about their assessment approaches to the public. Consequently, status quo assessment practices are largely opaque, with the best data available coming from surveys conducted by the International Association of Assessing Officers (IAAO). While these surveys are the best source of data for understanding the status quo, we caution that they are not representative.\looseness=-1} The remainder rely on a combination of \textit{valuation tables}, which estimate the marginal value of property attributes (e.g. additional square footage or bedrooms) and are  provided by states or purchased from private vendors; individual property assessments using comparables selected by hand; or other methods. The low adoption rate of AVMs has several causes. Roughly one in five IAAO survey respondents point to  capacity concerns, such as budget constraints or insufficient technical expertise. One in four cite difficulty explaining AVM predictions to taxpayers. However, the most-cited barrier to adoption is the belief that current approaches work just as well as AVMs, if not better, despite considerable empirical evidence to the contrary~\cite{bidansetSurveyUseAutomated2022}.\looseness=-1

Furthermore, according to a 2018 survey conducted by the IAAO, only one-third to one-half of local assessors have incorporated fine-grained location or neighborhood features into their assessment analyses~\cite{IAAO2018GISTechSurvey}. While traditional assessment models focus only on characteristics of a property itself (e.g., size or age), incorporating geographic data improves assessment by capturing location (e.g., proximity to schools) and neighborhood (e.g., sociodemographics of residents) information that is likely to affect a property's market value~\cite{droj2024comprehensive}.\looseness=-1

\subsection{The Fairness-Accuracy Tradeoff}
Fair ML research often casts training fair models as a constrained optimization problem: the goal is to maximize \textit{accuracy}, usually operationalized as predictive error, subject to a \textit{fairness} constraint, usually operationalized as differences in model error rates between groups~\cite{kamiran_classifying_2009, kamiran_data_2012, feldman_certifying_2015, chouldechova_fair_2016, kleinberg_inherent_2017, hardt_equality_2016, zafar_fairness_2017, corbett-davies_algorithmic_2017, menon_cost_2018, zhao_inherent_2022, berk_convex_2017}. This framing presumes the existence of a fairness-accuracy tradeoff, except under unrealistic circumstances such as perfectly accurate models~\cite{kleinberg_inherent_2017, pleiss_fairness_2017}. Because perfect accuracy is unachievable in practice, even the best model produces incorrect predictions for some instances. In theory, a tradeoff is inevitable if the most accurate model produces different errors across groups: if accuracy cannot be increased for the worst-off group, then the only way to achieve fairness is to reduce accuracy for the better-off groups. However, counter to the prevailing assumption in the fair ML literature that accuracy and fairness are mutually incompatible, an emerging strain of research has found that tradeoffs do not always arise in practice~\cite{cooper_emergent_2021}. For example, researchers have achieved simultaneous fairness and accuracy gains across a variety of modeling tasks by collecting additional data~\cite{chen_why_2018} and by accounting for bias in existing data~\cite{wick_unlocking_2019, dutta_tradoff_2020}.\looseness=-1

\subsection{Conceptual Framework}
In theory, accuracy gains in property tax assessment guarantee neither fairness gains nor fairness losses. For example, correcting errors in over-assessed low-value properties will generally reduce regressivity, while correcting errors in under-assessed low-value properties will increase it.

For at least one pair of accuracy and fairness metrics, the relationship between accuracy and fairness depends on the relative magnitudes of prediction variance and mean bias in a given assessment model. Formally, for two assessment models ($A$ and $B$), if accuracy is measured using mean squared error (MSE) between ground truth sale price and predicted sale price, and fairness according to the Log Coefficient regressivity metric described in Table ~\ref{tab:regressivity_metrics}, then: 
$$
    Fairness_B - Fairness_A  = \frac{1}{2} \left[ ( Var_B -  Var_A) + [(MeanBias_A^2 - MeanBias_B^2)] - [ MSE_B -  MSE_A] \right]
$$

\noindent To see that the relationship between fairness and accuracy is ambiguous, assume that models $A$ and $B$ have the same mean prediction ($MeanBias_A = MeanBias_B$) and that model $B$ has lower MSE ($MSE_B<MSE_A$). Then, if model $B$'s predictions are more variable than model $A$'s, accuracy and fairness will improve simultaneously. If model $B$'s predictions are less variable than model $A$'s then, depending on the magnitude of the difference in accuracy compared to the magnitude of the difference in variance, there may be a fairness-accuracy tradeoff. A formal proof is provided in Appendix~\ref{app:tradeoff}.\looseness=-1

As with other domains, a perfectly accurate model\textemdash one that sets the assessed value of each home equal to its market value\textemdash is also perfectly fair, as it produces a proportional property tax. However, whether imperfect assessment models eventually reach a Pareto frontier wherein accuracy gains can only be achieved with corresponding decreases in fairness is an open question. \textit{In this work, we explore empirically whether the fairness-accuracy tradeoff exists for property tax assessment, examining whether the types of accuracy improvements that are feasible in practice improve or undermine fairness.}\looseness=-1
\section{Data and Methods}\label{s-3-methods}

\subsection{Data}
Our data include property and transfer records purchased from the data broker Cotality.\footnote{Cotality was formerly called CoreLogic, and is widely used in economics research for research on property taxes and property transfers.} The unit of observation is a property transfer, which we filter to include only arms-length sales of single-family homes. Consistent with common practice, we exclude sales of distressed or foreclosed properties under the assumption that foreclosure auctions do not reveal a property's underlying market value. For each sale, we observe transaction data (e.g., type of deed, date of sale, sale amount). We then merge sale data with property characteristics (e.g., square footage, number of bedrooms, year built), location data (e.g., county, census block group), and tax information (e.g., assessed value at time of sale). Finally, we merge in publicly available socioeconomic data from the American Community Survey (ACS) 5-year estimates. These data include block group-level neighborhood information (e.g., education levels, marital status, median age and income, unemployment rates, and share of households using SNAP). Overall, this results in a dataset of 26M sales spanning the period 2018 to 2023 and covering transactions in 2,844 out of 3,007 U.S. counties.\looseness=-1

\subsection{Counterfactual Assessment Models}
Using the data described above, we build a suite of LASSO and random forest-based counterfactual AVMs. These models use property and neighborhood data to predict property sale prices, simulating the assessment process. Models are developed at the county level, using only within-county sales. We employ a chronological train-test split, such that models are trained on historical sales data and tested using the most recent year of sales, in order to align with real-world assessment practices~\citep[e.g.,][]{ccao_github_model}. To preprocess the data, we first one-hot encode categorical features, retaining only categories appearing in at least 5\% of observations. We then drop features whose values are missing for more than 50\% of observations, and use multiple imputation by chained equations (MICE)~\cite{mice} to impute any remaining missing values. Finally, we winsorize all numeric features at the 1st and 99th percentiles and subsequently normalize them. We conduct hyperparameter tuning using Bayesian optimization and five-fold cross validation (CV), with mean absolute error between predicted and actual log sale price determining the CV score.\looseness=-1 

%In addition to counterfactual models, we create baseline assessment estimates that are based on the actual assessed values of properties. This is necessary because in many counties, assessments are not intended to be estimates of market value at time of sale; rather, they are estimates of the market value at a hypothetical sale date that is held fixed across properties, reduced by some fixed percentage. For example, a county may set the assessed value of a home to be 10\% of its estimated market value. To account for this, we construct a baseline model for each county in each year using the same procedure described above but using only one explanatory variable: actual assessed values. This allows us to use existing assessments as a baseline while still conducting an apples-to-apples comparison that assumes that assessments are always seeking to approximate a property's sale price.\looseness=-1

\subsection{Evaluating Assessment Models}
We evaluate each assessment model in terms of its accuracy and fairness, as measured on the held-out most recent year of sales and assessment data for each county.\looseness=-1

\textit{Accuracy.} An accurate assessment is one that matches the true market value of a property, which is assumed to be equal to the sale price of a property in an arms-length transaction. We therefore measure accuracy as the difference between a property's assessed value in a given year (as produced by an assessment model) and its actual sale price in that year. As our primary accuracy metric, we use mean absolute percentage error (MAPE), because the percentage error between assessed and actual sale price is equivalent to the percentage difference between effective and statutory tax rates in a simple setting with no exemptions or deductions to the tax (see Appendix \ref{app:accuracy_etr}). To ensure robustness, we also evaluate model accuracy using root mean squared error (RMSE) and mean absolute error (MAE).\looseness=-1

%\paragraph{Challenges to Measuring Accuracy} 
%Following prior work, our approach to measuring accuracy assumes that a property's true market value is captured by its sale price. However, there is evidence that this may not be the case\textemdash specifically, sale prices may be subject to random noise, and the sales prices of homes sold by Black owners might be systematically lower than the true market values of those homes \cite{perryDevaluationAssetsBlack2018}. As far as we are aware, all existing approaches to measuring regressivity assume that a property's true market value is equal to its sale price, so in order to be consistent with prior literature and ensure that our accuracy and fairness measurements are meaningfully related to one another, we reproduce this assumption in our work. However, we discuss the implications of this assumption in detail in \S\ref{s-5-discussion}.\looseness=-1

\textit{Fairness.} We define fairness as the level of regressivity produced under a particular assessment model. Regressivity occurs when a higher proportion of the tax burden falls on the owners of lower-valued properties. It is commonly used within the field of economics to measure the vertical equity of tax systems \cite{black_algorithmic_2022, amornsiripanitchWhyAreResidential2022}.{We also conduct a more typical group parity assessment which compares how aggregate levels of regressivity change as the racial demographics or income of taxing jurisdictions change in \S\ref{subsec:censusraceincome}. 

\begin{table}[!t]
\centering
\begin{small}
\begin{tabular}{ m{2cm} m{1.7cm} m{7cm} m{1.5cm} m{1cm} }
\toprule
\textbf{Metric} & 
\textbf{Approach} & 
\textbf{Formula} & 
\textbf{Regressivity Threshold} &
\textbf{Metric Source} \\ 
\midrule
Log Coefficient (LC) & 
Regression-Based & $\beta_1$, where $\log(R) = \beta_0+\beta_1\log(S)$ and $R=\frac{A}{S}$. &
$<0$\newline (significantly) & \cite{berryReassessingPropertyTax2021, amornsiripanitchWhyAreResidential2022} \\
\midrule
Suits Index (SI) & 
Distribution-Based &  $1 - \frac{L}{K}$, where $L$ is the area under the curve produced by graphing the cumulative distribution of A against the cumulative distribution of S and $K$ is the area under an analogous curve produced by a perfectly proportional tax & 
$<0$\newline (significantly) & 
\cite{suits1977measurement, mcmillenMeasuresVerticalInequality2023} \\
\midrule
Price-Related Differential (PRD) & 
Ratio-Based &  
$\frac{\bar{R}}{\bar{A}/\bar{S}}$, where $\bar{R}$ is the average $\frac{A}{S}$, $\bar{A}$ is the average A, and $\bar{S}$ is the average S & 
$> 1.03$\newline (strictly) & 
\cite{iaaostandards} \\
\bottomrule
\end{tabular}
\caption{The regressivity metrics used in our analysis. Each metric compares the sale prices (S) and assessed values (A) of properties within a given geographic area in a given year.}\label{tab:regressivity_metrics}
\end{small}
\vspace{-20pt}
\end{table}

Local governments, professional associations, and researchers have collectively promulgated several competing regressivity metrics~\cite{mcmillenAssessmentRegressivityProperty2020}. As we discuss in detail in \S\ref{s-5-discussion}, no metric is without shortcomings and individual metrics may not always agree with one another. Therefore, we rely on three separate regressivity metrics throughout this analysis which together cover the primary approaches currently in use. We describe each metric briefly in Table~\ref{tab:regressivity_metrics}, and in detail in Appendix \ref{app:reg_metrics}.\looseness=-1

\textit{Pareto Frontier Framework.} To evaluate model performance, we use the framework of the \textit{Pareto frontier}. In multi-objective optimization problems, it is often the case that objectives are in tension with one another and cannot be simultaneously maximized. The tradeoffs among the best-performing models trace out a frontier (Fig.~\ref{fig:pareto_frontier_a}). Outcomes along the frontier are \textit{Pareto optimal}, in the sense that no other outcome is possible which offers improvements with respect to both objectives. If no tradeoff exists between the two objectives, then the Pareto frontier collapses to a single point, corresponding to a unique model that strictly outperforms all other models (Fig.~\ref{fig:pareto_frontier_b}). Such a model constitutes a \textit{Pareto improvement} over previous models, as it offers simultaneous gains with respect to both objectives.\looseness=-1

\begin{figure}
\centering
\begin{subfigure}{0.3\textwidth}
    \includegraphics[width=\textwidth]{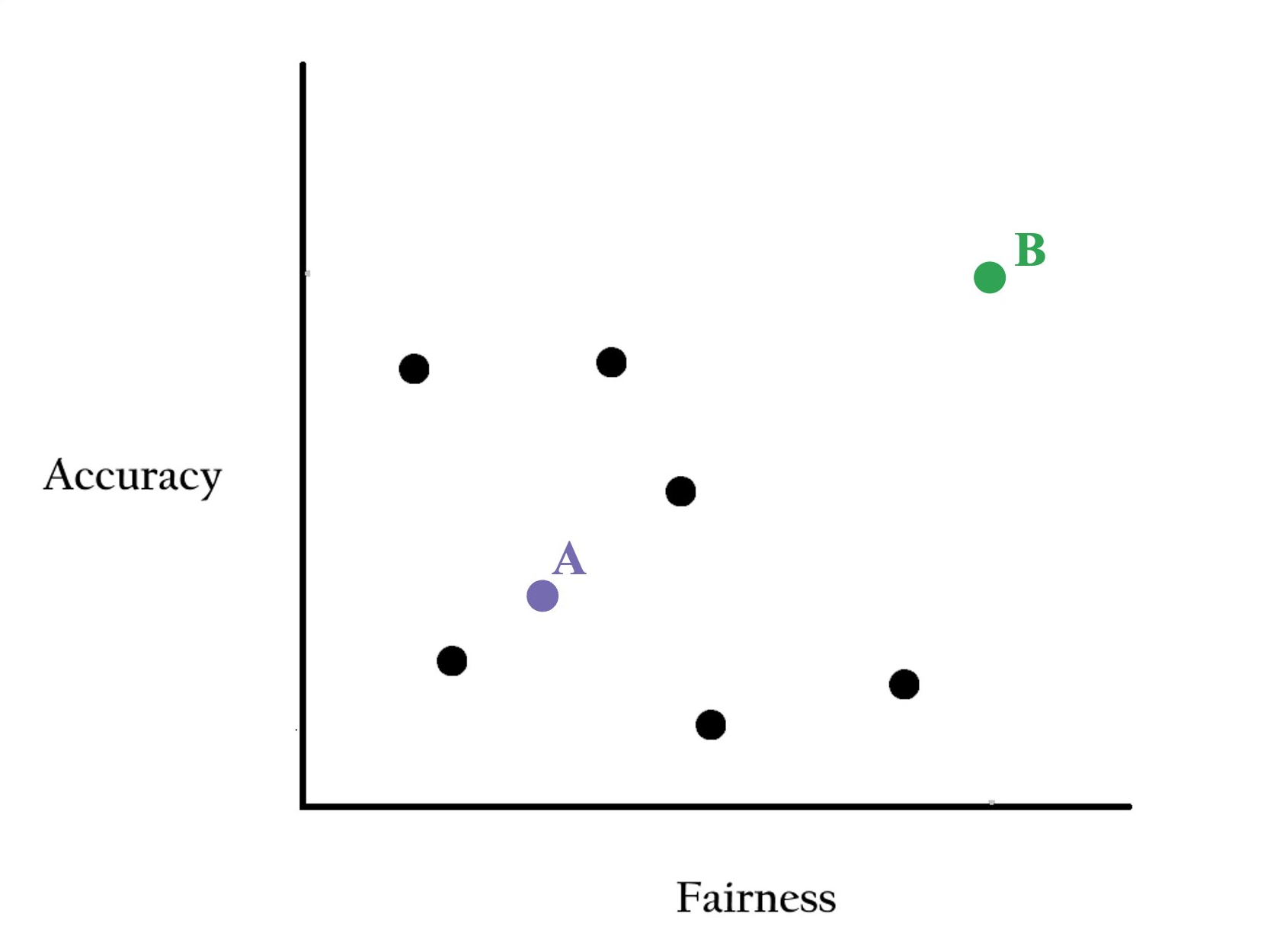}
    \caption{Current State}
    \label{fig:pareto_frontier_c}
\end{subfigure}
\hspace{0.03\textwidth}
\begin{subfigure}{0.3\textwidth}
    \includegraphics[width=\textwidth]{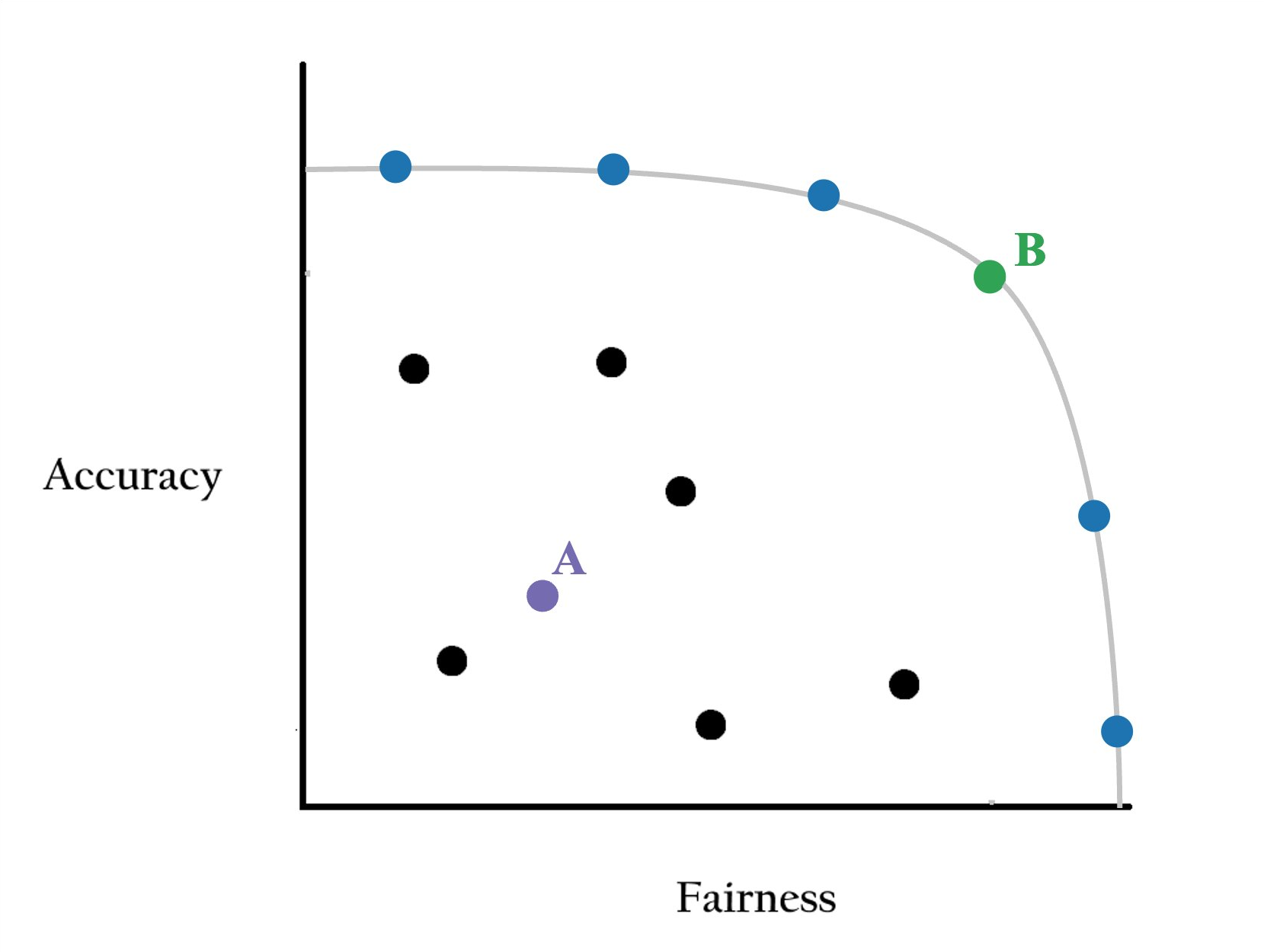}
    \caption{Pareto Frontier}
    \label{fig:pareto_frontier_a}
\end{subfigure}
\hspace{0.03\textwidth}
\begin{subfigure}{0.3\textwidth}
    \includegraphics[width=\textwidth]{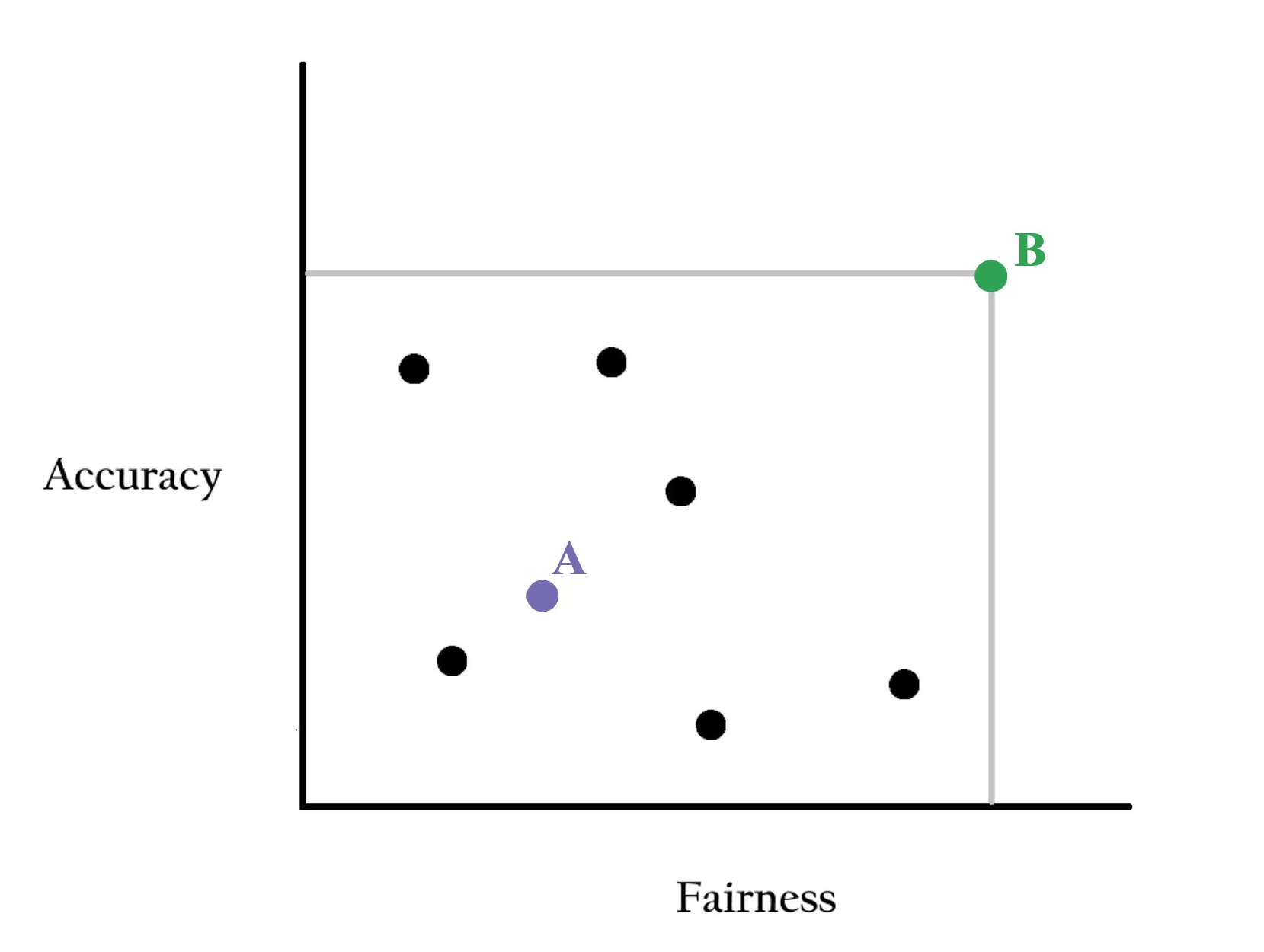}
    \caption{Pareto Point}
    \label{fig:pareto_frontier_b}
\end{subfigure}
\caption{We evaluate model performance using a Pareto framework. Each point represents a model. When model accuracy and fairness are low (point \textbf{A}), as is currently the case in property tax assessments, simultaneous improvements in accuracy and fairness (point \textbf{B}) are possible. This holds regardless of whether \textbf{B} represents a point on the Pareto frontier or is a uniquely best-performing Pareto point.\looseness=-1}\label{fig:pareto_frontier}
\Description{Three side-by-side graphs that provide a visual representation of our current state (leftmost graph), in which there are many different achievable models, some of which strictly dominate the others in terms of fairness and accuracy. In the current state, we do not know whether the Pareto dominant models exist on a Pareto curve (i.e., they cannot be further improved upon without a fairness-accuracy tradeoff) or whether there is a Pareto point (no fairness-accuracy tradeoff will ever exist). This is illustrated by a Pareto curve (middle graph) and a Pareto point (rightmost graph) superimposed over the current state.}
\end{figure}

Empirically, we examine whether counterfactual assessment models offer Pareto improvements\textemdash simultaneous gains in fairness and accuracy\textemdash relative to baseline predictions from either status quo assessments or other counterfactual models. If these modeling adjustments yield Pareto improvements relative to these baselines, then we posit that this is evidence against the practical importance of a fairness-accuracy tradeoff in property tax assessments.\footnote{Of course, the observed compatibility of fairness and accuracy gains relative to the baselines we consider does not imply that no such tradeoffs could ever occur; for example, they might arise if the status quo baseline were substantially improved (as depicted in Fig.~\ref{fig:pareto_frontier_a}). Our focus is on the practical importance of the fairness-accuracy trade-off in the context of the methods actually used to generate assessments in real-world settings.}
\section{Results}\label{s-4-results}

\enlargethispage{20pt}
In this section, we characterize the status quo relationship between accuracy and fairness in property tax assessment, finding that even though the relationship between accuracy and fairness is theoretically ambiguous, in practice the two measures are closely related. We first show that counties with more accurate status quo assessments also tend to have more equitable assessments (\S\ref{s-4.1-statusquo}). Next, we find that adding more information to hypothetical assessment models leads to accuracy gains in a majority of cases, and accuracy gains coincide with fairness gains over 99\% of the time. Finally, we show that, by incorporating publicly available census data in assessment models, simultaneous accuracy and fairness gains are possible relative to status-quo assessments in 14.4\% of U.S. counties without caps on annual assessment growth (\S\ref{s-4.2-improvements}).\looseness=-1

\subsection{The Accuracy-Fairness Relationship}\label{s-4.1-statusquo}

\subsubsection{Assessment Accuracy and Fairness are Highly Correlated, and Both are Low for Many Counties.}
\begin{figure*}[t!]
\centering  
\begin{subfigure}{0.32\textwidth}
        \includegraphics[width=\linewidth]{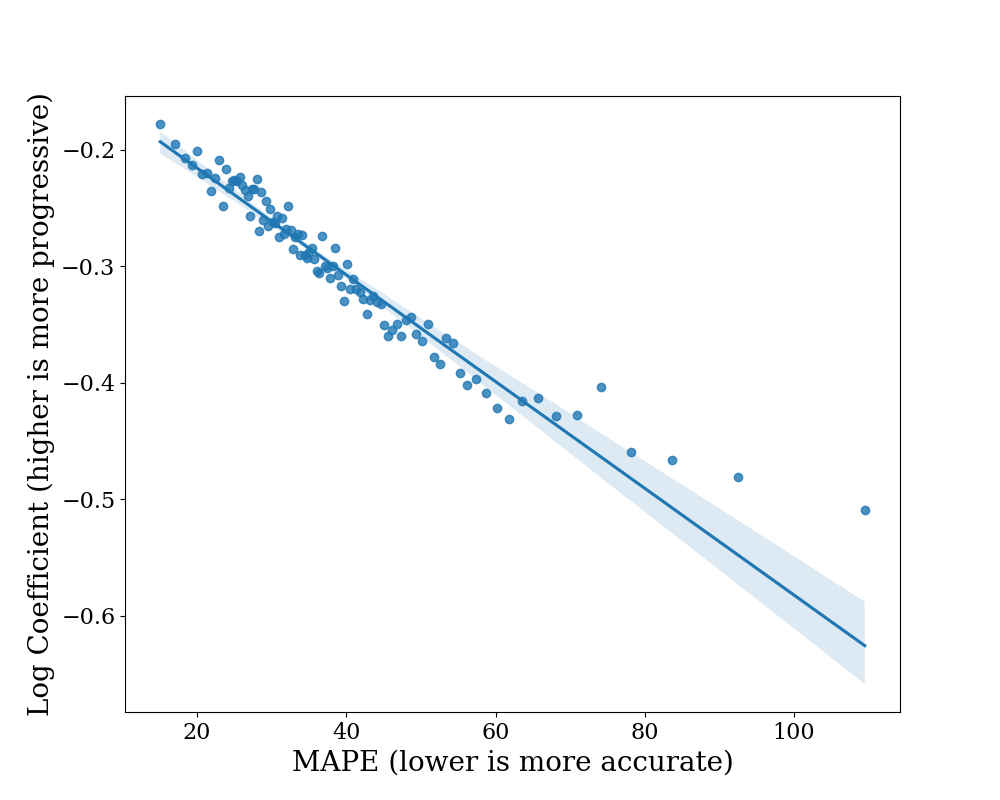}
        \label{fig:log_coef_sq}
        \caption{Log Coefficient}
    \end{subfigure}
 \hfill
    \begin{subfigure}{0.32\textwidth}
        \includegraphics[width=\linewidth]{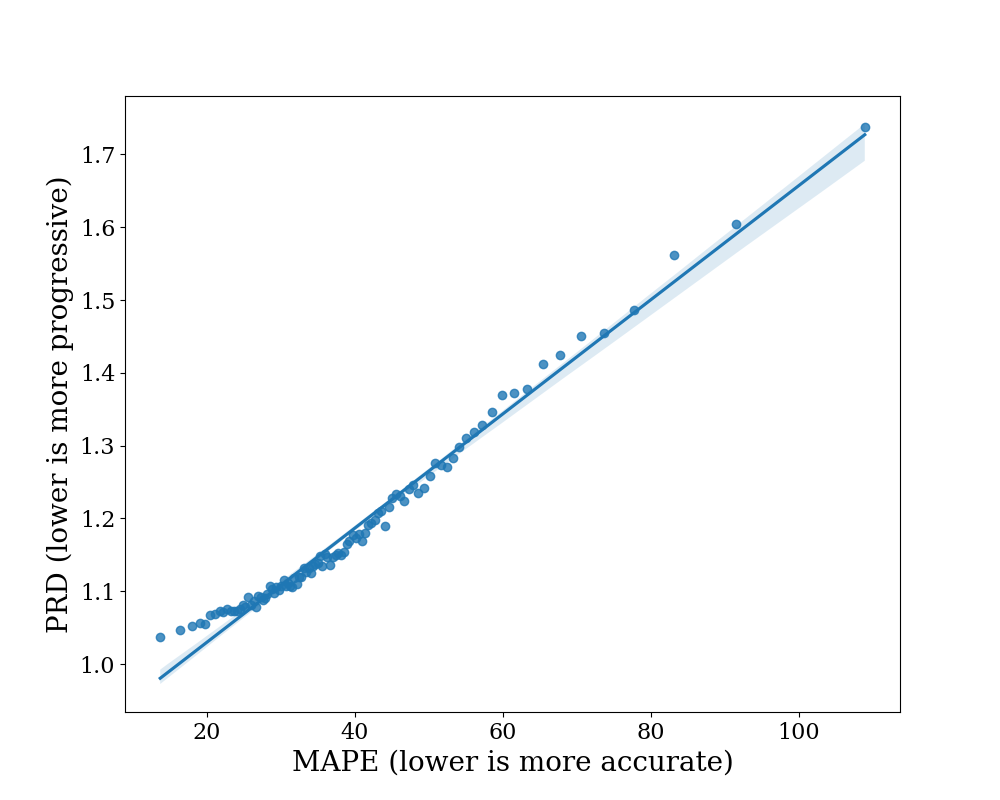}
        \label{fig:prd_sq}
        \caption{PRD}
    \end{subfigure}
\hfill
    \begin{subfigure}{0.32\textwidth}
        \includegraphics[width=\linewidth]{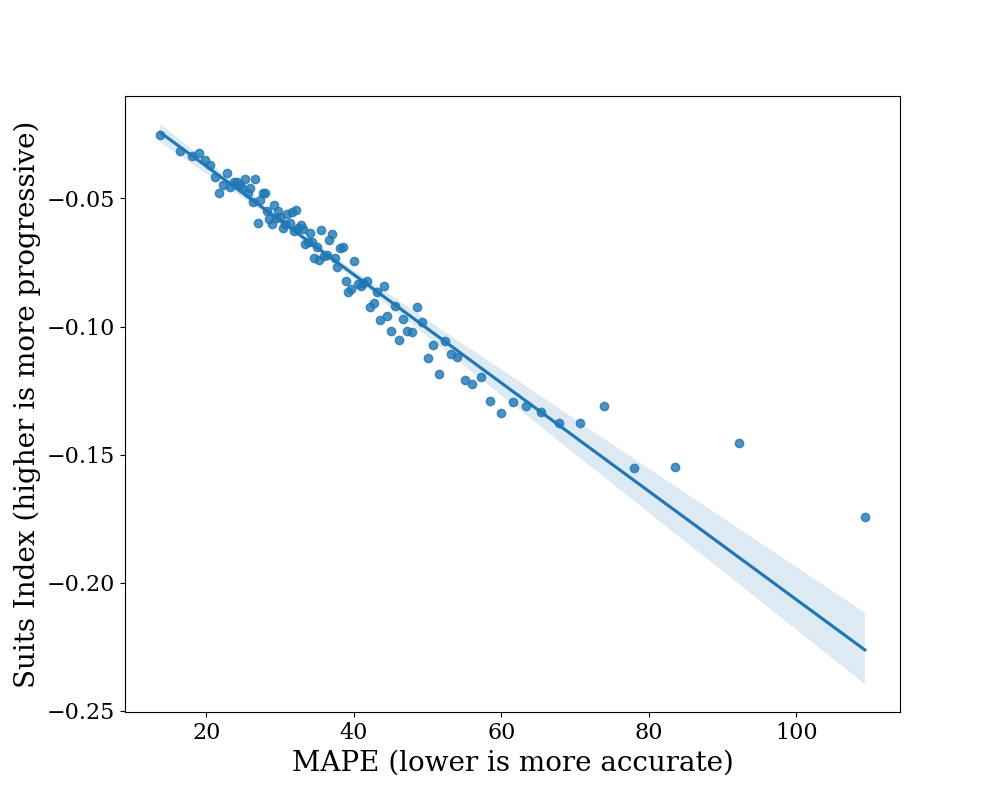}
        \label{fig:suits_sq}
        \caption{Suits Index}
    \end{subfigure}
\caption{Higher accuracy (as measured by MAPE) in property tax assessments is correlated with higher fairness. The figures show scatterplots of fairness and accuracy by county-year between 2018 and 2023, binned across 100 quantiles of county-level MAPE.}\label{fig:accuracy_fairness_status_quo}
\Description{Three scatterplots of fairness and accuracy by county-year between 2018 and 2023, binned across 100 quantiles of county-level MAPE. The figures show that as MAPE increases (gets less accurate), the Log Coefficient (leftmost graph) decreases (gets less fair), the PRD (middle graph) increases (gets less fair), and the Suits Index (rightmost graph) decreases (gets less fair).}
\end{figure*} 
As shown in Fig.~\ref{fig:accuracy_fairness_status_quo}, accuracy and fairness in property tax assessment are strongly correlated, and the relationship is robust across fairness metrics. In general, counties with a higher median and average sale price ($R=-0.23$ and $R=-0.17$, respectively), sale volume ($R=-0.19$), housing stock ($R=-0.18$), and population ($R=-0.17$) have lower status quo assessment errors ($p < 0.001$ in all cases). As a result, average county-level accuracy varies across states: states with a higher proportion of rural counties such as Oklahoma, Arkansas, and Missouri tend to have higher assessment error on average. Somewhat surprisingly, county-level variance in sale price is not associated with assessment accuracy, implying that more homogeneous housing markets do not, on average, produce higher-quality assessments. Almost all counties\textemdash 98\% between 2018 and 2023\textemdash had regressive assessments as measured by LC (i.e., LC $<0$).\looseness=-1

\subsubsection{Accuracy Gains Do Not Guarantee Fairness Gains.}
In our setting, fairness and accuracy are both functions of assessed value, and we observe a positive correlation between accuracy and fairness among status quo assessments. This raises the concern that assessment accuracy and fairness are not independent, since both are derived from assessed value. However, more accurate assessments are not necessarily fairer. As discussed in Section \ref{s-2-background} and shown in Appendix \ref{app:tradeoff}, the theoretical relationship between fairness and accuracy is ambiguous and hinges on changes in prediction variance between models. In this section, we consider empirical tradeoffs through the example of assessment appeals.\looseness=-1

Improvements in accuracy can increase regressivity if accuracy gains are primarily clustered among previously over-assessed high-value properties. This commonly occurs in the real world in the context of assessment appeals \cite{weber2010ask, mcmillen2013effect}. If a homeowner believes that their assessment is inaccurate, they can file an appeal with a county governing body. Appeals are disproportionately filed by owners of higher-valued properties, who have higher property tax bills at stake and are likely to have more resources available to navigate the appeals process. Because appeals typically reduce assessed values, the appeals system functions as an adversarial correction, increasing accuracy primarily for over-assessed, high-value properties while leaving other prediction errors untouched. The end result is that post-appeal assessments can be both more accurate and more regressive relative to pre-appeal assessments. The relationship between accuracy and fairness depends on the concentration of accuracy gains across the distribution of housing prices and on the direction of the errors that are corrected. We observe this empirically in Cook County, IL, where between 2021 and 2023, owners of homes in the highest quintile of sale price appealed their assessments at four times the rate of those in the lowest quintile~\cite{cook_county_treasurer_25}.\looseness=-1

\subsubsection{In hypothetical models, additional features do not impose fairness-accuracy tradeoffs in over 99\% of cases.}

To further examine the relationship between assessment accuracy and fairness, we develop simulated assessment models using property data from Cotality, examining whether models with access to more information about properties will be more accurate and equitable than models with less information. To test this, we build two LASSO models for each county according to the procedures described in \S\ref{s-3-methods}. We choose LASSO as opposed to more complex and compute-intensive approaches as it offers a simple framework suitable for establishing a baseline empirical relationship between data availability and fairness-accuracy tradeoffs. The first model (``sparse model") uses only three property characteristics. Because data availability varies across counties, these three features are selected from an ordered list of features ranked by their overall availability in Cotality. The most commonly available features\textemdash and therefore those most often included\textemdash are construction quality, land and building square footage, number of stories, and year built. The second model includes as many property characteristics as are available in Cotality for that county (``rich model"). Again, due to differences in data availability across counties, the additional number of features in the rich model relative to the sparse model ranges from 1 to 12 depending on the county. Both models include six features describing the timing of each sale. We subset to counties with at least 100 sales in our sample period (2018-2023), and to counties with a sufficient number of features available to train both sparse and rich models, producing a sample of 2,059 counties.\looseness=-1

Our results are provided in Fig.~\ref{fig:ablation_and_census_plots_alblation}. We observe significant changes in both accuracy and fairness in {908} out of 2,059 counties ({44.1}\%). In {907} of these {908} counties ({99.9}\%), we observe simultaneous gains in fairness and accuracy, while only 1 exhibits tradeoffs.\footnote{{DeSoto County, Missouri is the only county where we observe tradeoffs. This county is not an outlier with respect to demographics, data availability, housing characteristics, or publicly available data regarding housing assessment practices, and may represent a false positive given the scale of our experiment.}} Moreover, counties with larger accuracy gains also tended to have larger fairness gains. Appendix~\ref{app:robustness} shows the robustness of this result to alternative choices of accuracy and fairness metrics (Fig.~\ref{fig:ablation_plots_alt_metrics}).\looseness=-1

\begin{figure}[!t]
    \centering

    \begin{subfigure}{0.35\textwidth}
        \includegraphics[width=\linewidth]{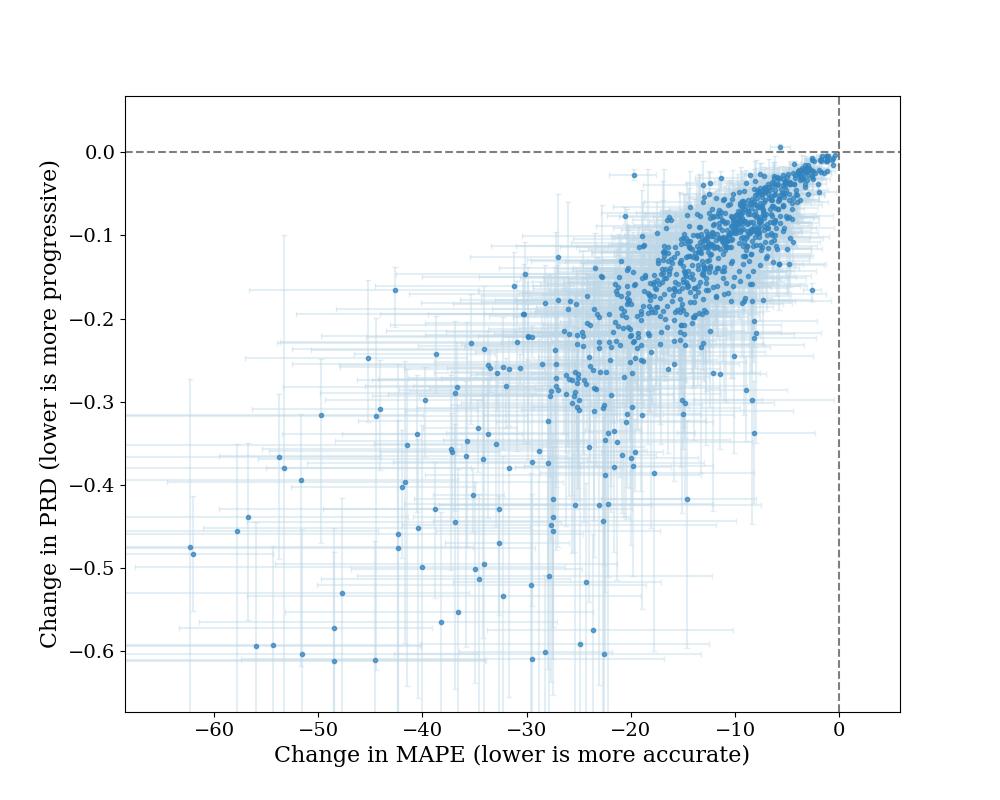}
        \caption{Ablation Experiment}\label{fig:ablation_and_census_plots_alblation}
    \end{subfigure}
\hspace{0.05\textwidth}
    \begin{subfigure}{0.35\textwidth}
        \includegraphics[width=\linewidth]{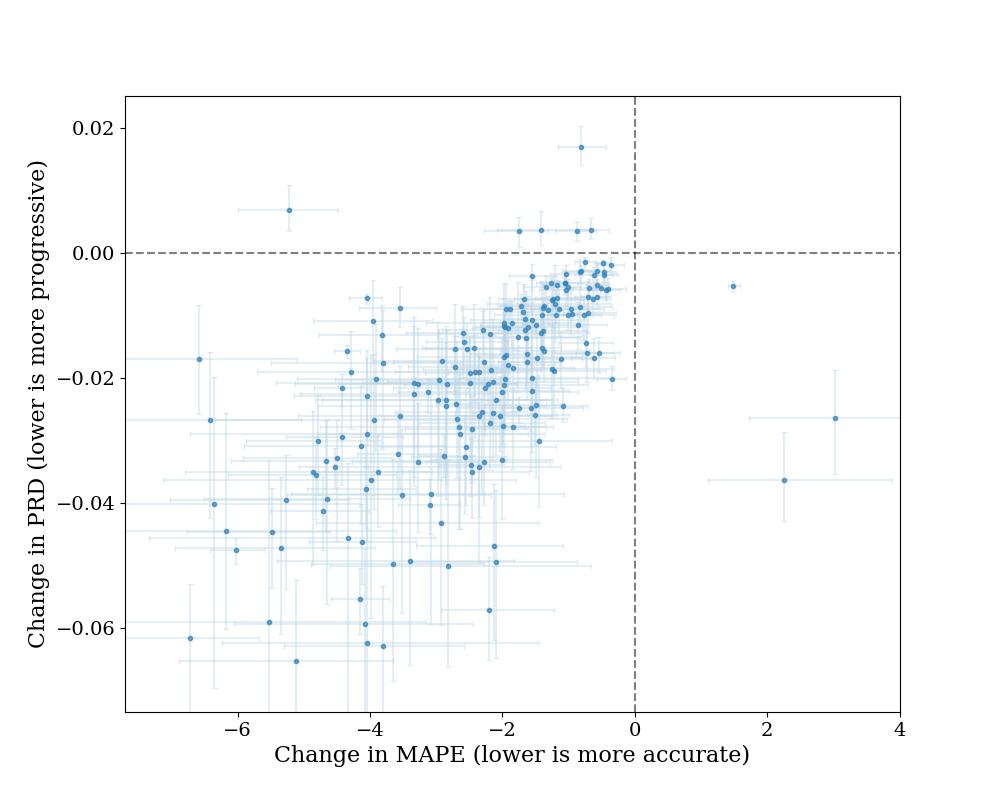}
        \caption{Census Experiment}\label{fig:ablation_and_census_plots_census}
    \end{subfigure}

    \caption{Adding more information to assessment models generally leads to accuracy and fairness gains.}
    \label{fig:ablation_and_census_plots}

    \begin{tablenotes}[flushleft]
    
    \scriptsize 
    \item \textit{Notes:} The figures show changes in accuracy and fairness for county-level assessment models across two experiments. The left panel shows the changes between the ``sparse" and ``rich" county-level LASSO-based assessment models described in Section \ref{s-4-results}. Models are trained using property and time-of-sale characteristics. The right panel shows  changes in accuracy and fairness after the addition of Census block group data to county-level random forest-based assessment models. The  baseline model uses only status quo predictions and time-of-sale data, while the alternative model incorporates these features in addition to Census block group characteristics. In both panels, accuracy is measured using MAPE between assessed and sale price. Fairness is measured using PRD.  Changes in performance are statistically significant at a 95\% confidence level after applying a Benjamini-Hochberg multiple testing correction. Confidence intervals are computed using a studentized bootstrap and displayed in light blue. Results are censored at the 2nd and 98th percentiles.\looseness=-1
    \end{tablenotes}
    \Description{The figures show changes in accuracy and fairness for county-level assessment models across two experiments: ablation and census. The left panel shows the changes between ``sparse" and ``rich" county-level LASSO-based assessment models. The right panel shows  changes in accuracy and fairness after the addition of Census block group data to county-level random forest-based assessment models. In both panels, accuracy is measured using MAPE between assessed and sale price. Fairness is measured using PRD. Both figures show a trend of simultaneously increasing fairness and accuracy.}
\end{figure}

The simple linear models developed here may be far enough below the Pareto frontier that the positive relationship we observe between fairness and accuracy is not sufficient evidence against fairness-accuracy tradeoffs in practical settings. To address this concern, in the next section, we rely on status quo assessments as our benchmark, and observe whether simultaneous gains in fairness and accuracy are possible relative to valuation methods currently in use.\looseness=-1

\subsection{Improving Property Tax Assessments Using Publicly Available Census Data}\label{s-4.2-improvements}

Housing prices are sensitive not only to the characteristics of homes, but also to the socioeconomic characteristics of the area surrounding a home, including local job opportunities and demographic characteristics such as age and household structure. The U.S. Census Bureau regularly disseminates estimates of these neighborhood-level characteristics, and at least one assessor's office uses such data in their valuation models.\footnote{Cook County, Illinois, whose assessment model documentation is publicly available: \href{https://github.com/ccao-data/model-res-avm?tab=readme-ov-file\#data-used}{github.com/ccao-data/model-res-avm}.} However, to our knowledge, this practice is not widespread, and guidance on the use of Census data in valuation models is not currently included in the technical standards promulgated by the IAAO \citep{IAAO2018AVM, IAAO2021DataQuality}. \citet{berryReassessingPropertyTax2021} identifies substantial within-county heterogeneity in assessment quality that is correlated with tract-level features such as median household income, education, and home values, while \citet{amornsiripanitchWhyAreResidential2022} finds that Census block-group characteristics improve assessment accuracy in aggregate across all counties. Both studies indicate that there is potential for these features to improve county-level model fit.\looseness=-1

We examine whether Census data {can improve the accuracy and fairness of assessment models currently in use} using publicly available data from the 2023 Census 5-year ACS, described in \S\ref{s-3-methods}. We train random forest models for each county that leverage Census data alongside status-quo assessed values and time-of-sale characteristics to predict sale price. {We use random forests because the Census features we employ are often correlated and may interact in complex ways. Random forests handle both properties well while remaining computationally tractable}. We then examine county-level changes in fairness and accuracy relative to predictions from benchmark random forest models which use only status-quo assessments and time-of-sale data. To account for statutory limitations which may impact the accuracy of status quo assessments, we drop counties located within 17 U.S. states with assessment caps \cite{dornfest2019state, dornfest2014state}.\footnote{These states include Alabama, Arizona, Arkansas, California, Florida, Georgia, Hawaii, Iowa, Louisiana, Maryland, Michigan, New Mexico, New York, Oklahoma, Oregon, South Carolina, and Texas. The annual caps range from 2 to 15\%.} We also restrict our sample to counties with at least 100 sales in our sample period, producing a sample of {1,619} counties.\looseness=-1

Our results are provided in Fig.~\ref{fig:ablation_and_census_plots_census}. We find that the addition of census features leads to statistically significant changes in fairness and accuracy for {238} out of {1,619} counties ({14.7}\%). Among those counties, {229} ({96.2}\%) show simultaneous fairness and accuracy gains, and only {9} ({3.8}\%) exhibit tradeoffs.\footnote{The counties for which we identified tradeoffs were: Washington, DC; Queens County, NY; Ashland County, OH; Geauga County, OH; Seneca County, OH; Obion county, TN; Robertson County, TN; King County, WA; and Manitowoc County, WI. There are some shared traits among these counties (e.g., some midsize Midwestern counties, some very large counties), but most counties with these traits do not exhibit tradeoffs. We do not find significant differences overall between counties with and without tradeoffs in terms of demographics, data availability, housing characteristics, or publicly available information about status quo assessment practices. Thus, while we can hypothesize about causes for tradeoffs, we also note the small magnitude of fairness-accuracy tradeoffs, which in some cases may capture noise as opposed to signal.\looseness=-1} The predominance of simultaneous fairness-accuracy improvements is robust to the choice of metric, as shown in Appendix \ref{app:robustness} (Fig.~\ref{fig:census_delta_alt}). As shown in Appendix Figure \ref{fig:quintile_comparison}, Census characteristics increase assessment accuracy on average across the entire distribution of sale prices, offering the most substantial gains for homes in the lowest and highest quintiles of price. Moreover, these characteristics tend to decrease assessed values for lower-priced homes and increase assessed values for higher-priced homes, which increases the overall variance of assessments and is consistent with fairness gains, as demonstrated in Appendix \ref{app:tradeoff}. \looseness=-1

As discussed in \S\ref{s-3-methods}, we use arms-length transactions to evaluate model performance. These sales represent a selected sample of the population of assessed properties, potentially favoring investment properties while excluding long-term owner occupants and foreclosures. To evaluate whether sample selection biases our results, we first compare key characteristics of sold and assessed properties in Appendix Table \ref{tab:sold_assessed_summ_stats}, finding no significant differences. Next, we select 10 counties at random from among the counties where we observe significant changes in fairness and accuracy and examine whether using sampling weights during model training and evaluation substantially changes our results. We define each property's sample inclusion probability as the share of single-family homes in that property's block group that sell in a given year, and we incorporate these probabilities as inverse probability weights. As shown in Appendix Table \ref{tab:census_weighted_prd_mape}, weighting does not substantially change our findings: signs are preserved in all but one case, and significance is preserved for 80\% of estimates.\looseness=-1

\subsection{Census Data Also Improves Outcomes By Race and Income}\label{subsec:censusraceincome}

While the predominant focus of this paper is regressivity with respect to sale price, there are other policy-relevant dimensions of fairness to consider, such as race and income. We find that incorporating Census characteristics in assessment models can also reduce regressivity with respect to block group-level measures of these characteristics. However, because we do not have access to individual-level data on homeowner race and income, and because our results are instead based on aggregate measures, they may suffer from aggregation bias and could reflect changes in outcomes for non-Black or wealthy residents in predominantly Black or poor neighborhoods. As such, these results are not conclusive regarding individual-level outcomes, but are rather intended to provide a basis for future work.\looseness=-1

\subsubsection{Race.}
\enlargethispage{20pt}
At the block group level, the addition of Census features in assessments has substantial effects on neighborhoods with relatively high shares of Black residents.\footnote{The Census features used here and throughout the paper do not include race or ethnicity.} As shown in Fig.~\ref{fig:av_mape_pct_blk}, Census block groups with more than $\sim$30\% Black residents see decreases in assessed value, on average, after the addition of Census features, while block groups with fewer Black residents tend to see increases in assessed value. Recall that because assessments are proportional to tax obligations, these results indicate that predominantly Black neighborhoods would see reductions in tax obligations, on average, from the addition of Census features in assessment models. Census features lead to across-the-board gains in accuracy, with slightly larger gains in percentage terms for predominantly Black neighborhoods.\looseness=-1

\begin{figure}[!t]
\centering
    \begin{subfigure}{0.35\textwidth}
        \includegraphics[width=\textwidth]{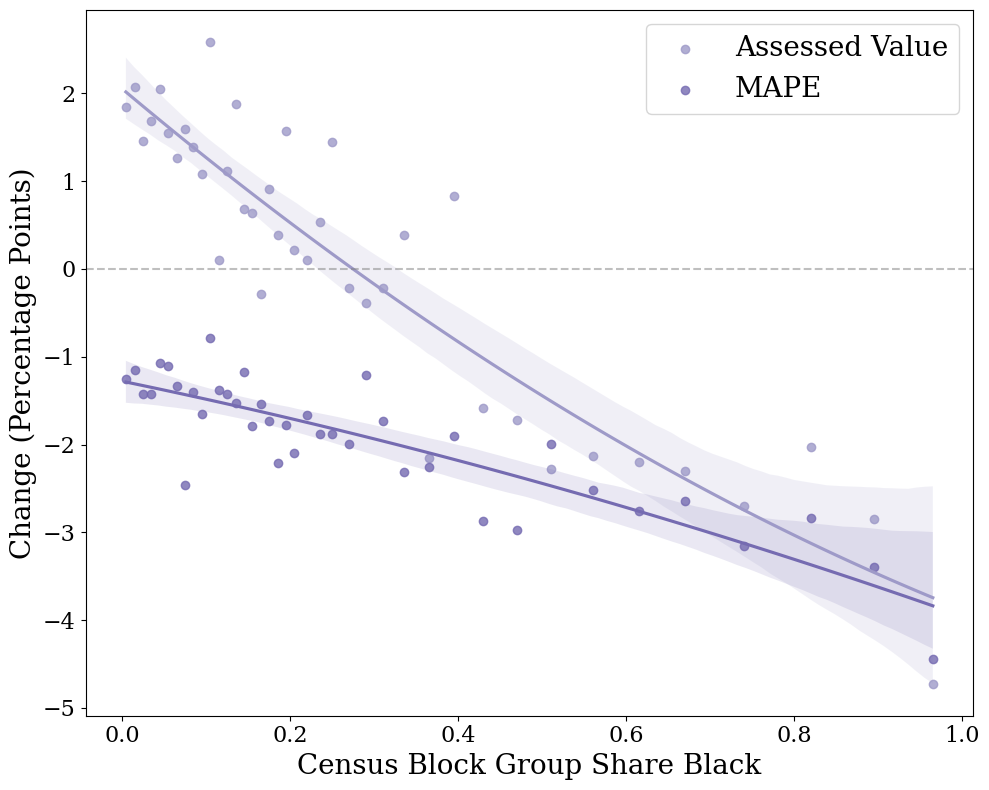}
        \caption{Race}\label{fig:av_mape_pct_blk}
    \end{subfigure}
\hspace{0.05\textwidth}
    \begin{subfigure}{0.35\textwidth}
        \includegraphics[width=\textwidth]{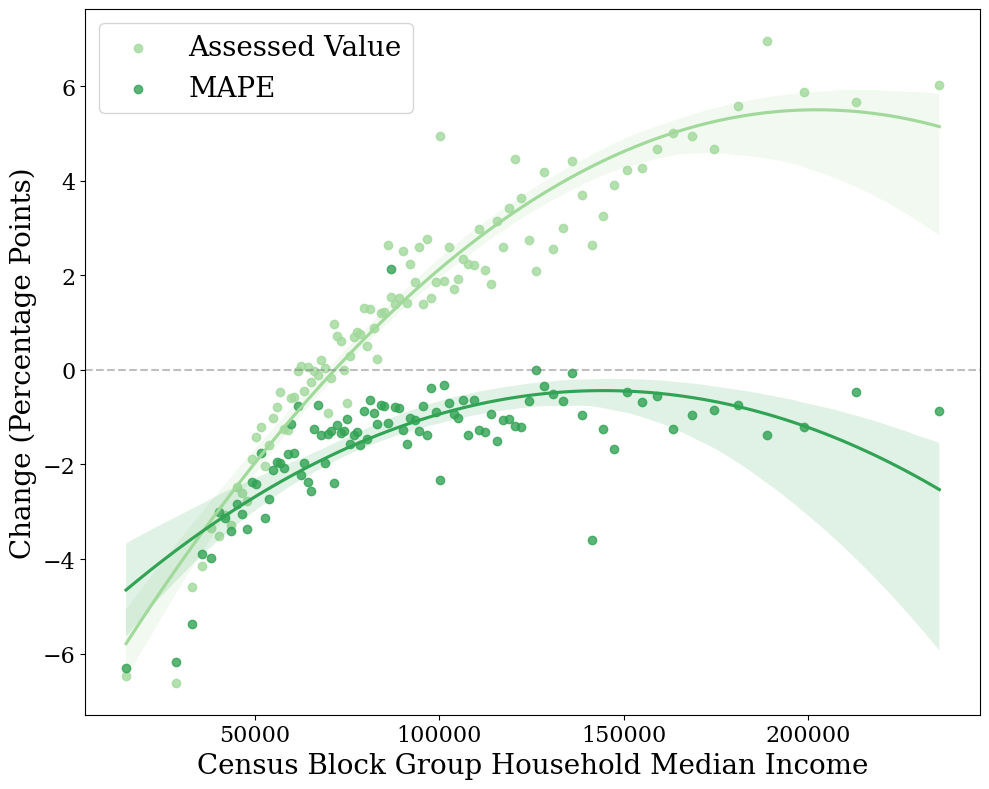}
        \caption{Income}\label{fig:av_mape_med_inc}
    \end{subfigure}
        
\caption{Census features lower assessed values for predominantly Black and low-income neighborhoods and increase accuracy overall.}
\label{fig:metric_changes_by_cbg_chars}

\begin{tablenotes}[flushleft]
    
    \scriptsize 
    \item \textit{Notes:} The figure shows changes in assessed value and MAPE relative to time-adjusted status-quo assessments after the addition of Census block group features in assessment models. Changes are calculated at the block group level, and observations are grouped by bins of the share of Census block group residents that are Black (a) and household median income (b). Lines of best fit are calculated using second-degree polynomials.\looseness=-1
    \end{tablenotes}
    \Description{The figure shows changes in assessed value and MAPE relative to time-adjusted status-quo assessments after the addition of Census block group features in assessment models. The left panel shows changes grouped by the share of a Census Block Group that is Black and shows that the inclusion of Census features lowers both assessed values and MAPE for predominantly Black neighborhoods. The right panel shows changes grouped by Census Block Group median household income and shows that the inclusion of Census features reduces MAPE.}
\end{figure}

\subsubsection{Income.}

With respect to income, as shown in Fig.~\ref{fig:av_mape_med_inc}, the addition of Census features tends to reduce assessments in lower-income neighborhoods and increase assessments in higher-income neighborhoods, with adjustments roughly proportional to income levels. Accuracy improves across the board, with the largest reductions in error in percentage terms concentrated among low-income neighborhoods. These results indicate that the addition of Census features could shift assessed values and corresponding property tax obligations up the income distribution, making the tax more progressive with respect to income as well as property values.\looseness=-1
\section{Discussion}\label{s-5-discussion}

\enlargethispage{20pt}
Our findings provide strong evidence against fairness-accuracy tradeoffs in property tax assessments relative to the current frontier of assessment models. While we show that tradeoffs are theoretically and empirically possible, we also find that the practical importance of these tradeoffs is limited. Rather, improvements in model accuracy tend to coincide with fairness gains, a conclusion which is robust across the vast majority of U.S. taxing jurisdictions and to various combinations of domain-relevant accuracy and fairness metrics. Moreover, we demonstrate that Pareto improvements are possible relative to the frontier of status quo models and predictions for a significant portion of counties in our data, and that these improvements can be achieved by incorporating publicly available Census data into assessment models. These findings lie in sharp contrast to other settings\textemdash predominantly, but not exclusively, involving classification\textemdash  where fairness-accuracy tradeoffs have been identified. In those settings, it is often the case that enforcing notions of fairness, such as statistical parity or equalized odds, necessitates reductions in predictive accuracy, either empirically or through the joint incompatibility of fairness and accuracy as modeling objectives \cite{chouldechova_fair_2016, compas, chouldechova_case_2018, quinonero_candela_disentangling_2023, koenecke_popular_2023, berk2017convex}.\looseness=-1 

\enlargethispage{6pt}
The key distinguishing features of our setting are, first, a focus on vertical equity (treating different individuals appropriately differently) as opposed to horizontal equity (treating similar individuals from different groups in similar ways), and second, our conception of fairness, which compares the distribution of assessment errors relative to ground-truth sale prices. In this framework, tradeoffs are theoretically possible, but they are not implied by the mutual incompatibility of fairness and accuracy, and we do not observe them empirically in the overwhelming majority of cases. This implies that our findings could generalize to other forms of wealth taxation beyond the property tax, such as taxes on inherited wealth and overall assets which are prevalent in countries outside the U.S. \cite{drometer2018wealth}. We also find preliminary evidence that the modeling improvements we propose here could reduce regressivity with respect to race and income.\looseness=-1 

\subsection{Limitations and Extensions}\label{subsec:limitations}
In this section, we discuss limitations of our analysis caused by modeling choices as well as by inherent challenges to measuring fairness and accuracy in property tax assessment at scale. Throughout, we highlight opportunities for future work.\looseness=-1

\textit{Scope of ``no tradeoff" claim.} In this paper, we provide comprehensive empirical evidence against fairness-accuracy tradeoffs relative to two baselines: a sparse linear model (ablation experiment), and status-quo assessments (Census experiment). We study the prevalence of tradeoffs using two supervised modeling frameworks (LASSO and random forest), and the primary experimental variation we employ is the addition of new data. While we consider status-quo assessments to be the appropriate and policy-relevant baseline for evaluating tradeoffs, and the incorporation of publicly available data to be a low-cost, feasible policy intervention, we recognize that, under different modeling assumptions and baselines, fairness-accuracy tradeoffs may be more prevalent. In particular, our findings may not generalize to settings where assessment agencies are already operating near the efficiency frontier, or where the scope for improvement is constrained by data availability, institutional capacity, or legal restrictions on the use of certain property characteristics.\looseness=-1

\enlargethispage{20pt}

\textit{Data Access and Representativeness.} Most counties do not publicly release data on property sales and assessments. As a result, we relied on brokered data from Cotality to conduct a national analysis. However, this brokered data contains a non-representative sample of single-family home sales when compared to county-level property records~\cite{harvey2025cotality}. As a result, while evidence of regressivity exists in both brokered data and county-level records, the levels of regressivity depend on the source. In addition, we chose to exclude counties with fewer than 100 sales during our sample period from our analysis (3.9\% of counties). We note that our suggestions to improve assessment quality may not be as effective in counties with lower sales volumes, as those counties have limited data to train assessment models. Future work might consider the possibility of data sharing between counties, and the relative tradeoffs, if any, associated with performing assessments at higher units of government, such as the state level, which may help alleviate data sparsity at the local level.\looseness=-1

More fundamentally, the standard approach for evaluating assessment quality (which we employ) is to compare assessed values to sale prices for the population of homes that sell in a given year. However, sold homes may differ in important ways from the overall population of assessed properties, and these differences can bias estimates of assessment quality. We take preliminary steps in the paper to estimate the impact of sample selection on our results, in particular by incorporating sampling weights for a subset of counties in our Census experiment, and we observe no significant impact on our findings. While we provide preliminary evidence that sample selection is unlikely to drive our conclusions, we acknowledge that a more comprehensive treatment of this issue remains an important direction for future work.\looseness=-1

\textit{Does Sale Price Represent the True Value of a Home?} Throughout this analysis, we have assumed that sale price represents the true value of a home. However, sale prices may deviate from true values for a number of reasons. For example, there may be idiosyncratic differences in negotiating skill across buyers and sellers, which can contribute to regressivity \cite{berryReassessingPropertyTax2021}. Sale price and true home values may also differ for more systematic reasons. Racial biases on the part of buyers, sellers, and other market actors, as well as historical legacies of redlining and disinvestment from non-white neighborhoods, can introduce racial disparities in sale prices that are unrelated to the observable characteristics of a home. Indeed, \citet{perryDevaluationAssetsBlack2018} find that, in appraisals conducted at time of sale, properties in majority-Black neighborhoods are devalued by as much as 23\% relative to comparable homes in majority non-Black neighborhoods, even after controlling for property and neighborhood characteristics. Race-based devaluation of homes both within and across neighborhoods is an important social problem with implications for wealth accumulation and intergenerational wealth transfer. To the extent that property tax assessments fail to account for race-based differences in home values, they can engender or worsen racial wealth gaps. In other words, if property tax assessments do not account for racial devaluation, then homeowners in majority-Black neighborhoods face a double penalty: when their homes are sold, they fetch a lower price; when their homes are taxed, it is at a higher rate. Concretely, this means that Black and Hispanic residents currently pay an estimated 10 and 13\% more in property taxes, respectively, for the same bundle of public services, after controlling for taxing jurisdiction, income, and property characteristics \cite{avenancio-leonAssessmentGapRacial2022}. Even though the sale price of homes in majority-Black neighborhoods should not necessarily be interpreted as the fair market value of those homes, bringing assessments closer to those sale prices would still improve fairness by reducing tax burdens in historically over-taxed, majority-Black neighborhoods.\footnote{Although we argue that reducing error between assessed value and sale price will make tax burdens fairer \textit{even when sale price is a biased estimate of fair market value}, this problem cannot be fully solved without fixing the systematic devaluation of homes in majority-Black neighborhoods. We discuss policies intended to address this issue in \S\ref{subsubsec:policyrecs}.}
Encouragingly, we find that the inclusion of Census block group characteristics in assessment models can identify a portion of these differences, increasing accuracy and lowering average assessments in aggregate in predominantly Black neighborhoods. This suggests that assessors can achieve Pareto improvements in accuracy and fairness in aggregate with respect not only to home values, but also neighborhood-level race, at little to no cost.\looseness=-1

Addressing what remains of the racial gap in property tax assessments after accounting for neighborhood-level features is a more challenging task. Property tax assessors could theoretically include homeowner race as an explicit feature in assessment models, recognizing the contextual value of race in predicting home prices. However, property tax assessors typically do not know individual homeowners' race, and would therefore be forced to rely on imputed measures of race, which introduce additional complications~\cite{elzayn_estimating_2024, Elzayn_measuring_2024}.
In addition, under prevailing anti-discrimination law, government entities may engage in race-conscious decisionmaking only if they can demonstrate a ``strong basis in evidence" that disparities exist, that they are personally accountable for creating them, and that corrective action is necessary and justified \cite{ho2020affirmative}. This is a high evidentiary standard. Recent court decisions have trended towards an interpretation of the law that would bar the explicit inclusion of race in assessment models regardless of the intended purpose. A more promising approach for interventions may therefore be to establish a firmer understanding of the underlying mechanisms driving bias in sale price and intervene directly at those levels. For example, if racial biases drive wedges between market value and sale price for homes sold by minorities, then jurisdictions might consider blinded sales. While the examination of accuracy-fairness tradeoffs in property tax assessments with respect to homeowner race is not the central focus of this paper, we provide evidence at the aggregate level which suggests that improving assessment accuracy can help reduce racial assessment disparities. Directions for future work include examining the impact of individual-level race and other demographic data on model performance, and assessing the legal and procedural challenges of including such data as explicit features of assessment models.\looseness=-1

\textit{No Standard Measure of Regressivity.} Throughout this paper, we have estimated regressivity by comparing assessed values to sale prices.  However, there is no agreed-upon `best' method for doing this---while the IAAO recommends relying on ratio-based approaches~\cite{iaaostandards}, economists have argued that those approaches can inflate estimates of regressivity~\cite{mcmillenMeasuresVerticalInequality2023}. Different regressivity metrics can disagree with one another on whether regressivity is present, further complicating measurement~\cite{mcmillenMeasuresVerticalInequality2023}. Furthermore, there are several other dimensions of regressivity that may be of interest to assessors and policymakers. For example, as an alternative to assessed values, one could examine property tax bills, which differ from assessments to the extent that taxpayers claim the credits and deductions offered by their taxing jurisdictions. In particular, many jurisdictions offer homestead exemptions to taxpayers who live in the property being taxed, as well as tax credits to seniors. Such exemptions can, but do not always, offset assessment regressivity \cite{ihlanfeldt2022homestead, mcmillenAssessmentRegressivityProperty2020}. One could also examine the regressivity of assessments or property tax bills relative to household income, as opposed to property sale price. This is the traditional approach that economists take when estimating the progressivity of a tax, and we provide preliminary evidence in this paper that reducing assessment regressivity with respect to sale price can also reduce statutory regressivity with respect to block-group level median incomes. Estimating tax regressivity with respect to individual-level incomes, however, is more complex, requiring the linking of administratively separated data sets and methodological decisions over the appropriate measure of income (annual versus lifetime), incidence (statutory versus economic), and conception of the property tax itself (capital tax versus fee for service) \cite{oates2016local}.\looseness=-1

\subsection{Recommendations}

We close by making several recommendations for the literature on accuracy-fairness tradeoffs and for tax assessors and policymakers seeking to improve the fairness and accuracy of property tax assessments.

\subsubsection{Recommendations for Studies of Fairness-Accuracy Tradeoffs.}

Our paper engages with two important critiques of the existing literature on fairness–accuracy tradeoffs: first, that the data and metrics employed in this work are too often disconnected from real-world applications and domain-specific conceptions of equity \cite{wagstaff2012machine}; and second, that the bias mitigation strategies found to impose tradeoffs are often ad hoc and intervene at just a single point in the modeling pipeline \cite{black2023toward}. We find that, in a concrete ML application of practical import, and with a holistic approach that spans the entire modeling pipeline and applies domain-relevant fairness metrics, tradeoffs are the exception rather than the rule, and that there exist substantial opportunities for Pareto improvements relative to the status quo.\looseness=-1

Our findings do not negate the statistical and theoretical validity of fairness–accuracy tradeoffs identified in previous work. Indeed, tradeoffs may arise in assessments under alternative modeling choices, baselines, or fairness definitions that abstract away from how domain experts have historically conceptualized equity. However, our results suggest that the practical relevance of these tradeoffs depends critically on examining the full modeling pipeline, using domain-appropriate metrics, and evaluating performance relative to models currently in use. Greater attention to domain-relevant notions of fairness can help the fair ML literature more accurately identify when tradeoffs have real-world significance.\looseness=-1

\subsubsection{Recommendations for Assessment Methodology} 
\leavevmode\vspace{2pt}

\textit{Collect More High-Quality Data on Properties.} One of the fundamental challenges of property tax assessment is that assessors often have less information about properties than buyers and sellers. Assessors typically cannot see the interiors of properties and instead rely only on information they can gather from properties' exteriors and from permitting records. We find that adding more information to assessment models generally leads to Pareto improvements in model performance. While there are concrete limits to the amount of information assessors can collect, there are likely also opportunities for many local governments to improve the coverage and quality of the data at assessors' disposal. Our findings indicate that more frequent and comprehensive field surveys, alongside investments in data validation, would in most cases lead to to better and more equitable assessments. However, not every county may be in a position to invest additional resources into collecting and maintaining more current and comprehensive data.\looseness=-1

\textit{Incorporate Census Block Group Characteristics.} We find that roughly 14.4\% of counties in states without assessment caps could improve the accuracy and uniformity of their assessments by adding Census block group features into their assessment models. This data is high quality, publicly available, and can be incorporated into assessment pipelines at little to no cost. While counties should independently evaluate the impact of Census data on model performance, in many cases this data is likely low-hanging fruit for counties looking to improve assessments.\looseness=-1

\textit{Use Tree-Based Automated Valuation Models.} As discussed in Section~\ref{s-2-background}, only a small minority of assessor's offices have adopted AVMs~\cite{bidansetSurveyUseAutomated2022}, despite a wealth of evidence indicating that AVMs outperform simpler assessment methods~\cite{bidanset2017accounting, droj2024comprehensive, jennifer2021garbage, quintos2014improving}. Moreover, among offices that have adopted AVMs, linear regressions remain the model of choice, even though tree-based models generally outperform linear models in predicting housing prices~\cite{fan2006determinants, yilmazer2020mass, hong2020house, berry2025evaluation}. In the real world, counties that have adopted tree-based AVMs have seen noteworthy accuracy and fairness improvements. For example, Cook County, Illinois, which publicly overhauled its assessment process after a high-profile investigation documented widespread property tax regressivity~\cite{cst_fbi, ct_regressivity, ct_regressivity2, ct_regressivity3, ct_regressivity4}, has used a LightGBM-based assessment model since 2019. In that time, Cook County has reduced property tax obligations by over one billion dollars for homeowners in the bottom 70\% of the housing price distribution, and has also come into compliance with industry standards regarding assessment uniformity \cite{berry2025evaluation}. We recommend that assessor's offices consider adopting AVMs, and tree-based AVMs in particular.\looseness=-1

\textit{Release Open Data on Sales and Assessments.} Publicly available data on sales and assessments, alongside transparent documentation of assessment methodology, would provide all stakeholders\textemdash policymakers, academic researchers, and ordinary citizens alike\textemdash the opportunity to understand how property tax bills are determined and how assessments might be improved. More transparency surrounding county-level assessment practices could also help paint a more thorough picture of the current frontier of assessment practices and assist counties in learning from one another about new approaches to assessment. Moreover, open data would enable researchers to validate the coverage and accuracy of brokered datasets, which form the basis for much of the existing research on property tax assessments. Cook County, which publicly releases its assessment data and modeling details, can serve as a template here.\footnote{\href{https://github.com/ccao-data/model-res-avm?tab=readme-ov-file\#data-used}{github.com/ccao-data/model-res-avm}}\looseness=-1 

\subsubsection{Recommendations for Assessment Policy}\label{subsubsec:policyrecs}
\leavevmode\vspace{2pt}

\textit{Consider Implementing Book-Tax Conformity for Assessments.} ``Book-tax conformity" refers to the concept that corporations should report the same income for tax purposes that they report on their financial statements, with the idea that doing so will encourage firms to reduce tax avoidance \cite{hanlon2005book}. The analog in the property tax setting would be conformity between assessed values and appraised values/sale prices. Homeowners would prefer that their assessed values remain as low as possible to minimize their property tax bills, but may prefer to have higher appraised values and sale prices to maximize their proceeds from the sale of a home, or to secure home equity loans. Aligning assessed values with appraisals and sale prices would exploit this tension and eliminate regressivity for homes that sell. It could also reduce racial disparities in regressivity: Black-owned homes have been shown to be consistently under-appraised at time of sale~\cite{perryDevaluationAssetsBlack2018} even as they are over-assessed in the property tax system~\cite{avenancio-leonAssessmentGapRacial2022}, as discussed in detail in \S\ref{subsec:limitations}. Of course, these advantages apply only to homes that are appraised or that sell. Moreover, book-tax conformity could cause discrete jumps in tax obligations upon sale that introduce frictions in the housing market. Ultimately, the potential benefits of book-tax conformity in assessments are context dependent.\looseness=-1
\enlargethispage{20pt}
\textit{Reconsider Statutory Limits on Assessments.} A majority of states regulate the way assessments are conducted and the impact of assessments on property tax bills. In particular, 17 U.S. states impose caps on the year-over-year growth in assessed values \cite{dornfest2019state, wasi2005property, dornfest2014state, beebe2025texas}. These caps limit the impact that improved assessment methodology can have on property tax regressivity. While more accurate predictions can lower assessed values and tax bills for over-assessed (and disproportionately low-value) properties, assessment caps constrain the extent to which these tax obligations can be shifted to under-assessed (and disproportionately high-value) properties, undermining the impact of assessment methodology on vertical equity. Additionally, 15 U.S. states only require counties to re-assess properties every five to six years, while an additional 10 states enact no requirements at all on the frequency of re-assessments \cite{dornfest2019state}. Even if initial assessments are equitable, differences in market dynamics across neighborhoods and property types can drive differential wedges between assessed and market value over time \cite{berryReassessingPropertyTax2021}. As such, whether an assessment model is accurate or not matters less and less as the period between assessments grows. While improvements in assessment methodology are a promising avenue for improving the vertical equity of the property tax in many jurisdictions, state-level regulations on assessments will inevitably mediate how much modeling improvements can reduce assessment regressivity.\looseness=-1
\section{Conclusion}\label{s-6-conclusion}

This paper examines whether property tax assessments are subject to fairness-accuracy tradeoffs, and finds that good-faith efforts to increase assessment accuracy generally lead to fairer, more uniform assessments, even though tradeoffs are both theoretically and empirically possible. In particular, adding Census block group characteristics to assessment models could significantly increase both assessment accuracy and fairness, with accuracy and fairness gains coinciding in {96}\% of cases. This finding stands in stark contrast to many other settings where fairness-accuracy tradeoffs have been examined, and contributes to a small but growing literature examining tradeoffs in regression settings. We provide recommendations for policymakers and assessors and suggest directions for future work, including examining other dimensions of fairness and addressing data and measurement challenges in the assessment space. Fundamentally, our work challenges the assumption that fairness and accuracy are necessarily in tension, and shows that machine learning can jointly advance both objectives, revealing a wider range of domains where fairness comes for free.

\begin{acks}
In addition to our referees, we thank the participants of the Stanford RegLab Roundtable, the Syracuse-Chicago Webinar Series on Property Tax Administration and Design, and the National Tax Association's 2025 Annual Meeting\textemdash in particular our discussant Iuliia Shybalkina\textemdash for their helpful comments, which greatly improved the paper. We also thank Myer Blank for helpful conversations.
\end{acks}

\bibliographystyle{ACM-Reference-Format}
\bibliography{references}

@misc{proptaxhistory,
title={History of Property Taxes in the United States}, 
author={Glenn Fisher},
editor={Robert Whaples},
year={2002}, 
month={09},
url={https://eh.net/encyclopedia/history-of-property-taxes-in-the-united-states/}, 
howpublished={EH.Net Encyclopedia}
}

@InProceedings{dutta_tradoff_2020,
  title = 	 {Is There a Trade-Off Between Fairness and Accuracy? {A} Perspective Using Mismatched Hypothesis Testing},
  author =       {Dutta, Sanghamitra and Wei, Dennis and Yueksel, Hazar and Chen, Pin-Yu and Liu, Sijia and Varshney, Kush},
  booktitle = 	 {Proceedings of the 37th International Conference on Machine Learning},
  pages = 	 {2803--2813},
  year = 	 {2020},
  editor = 	 {III, Hal Daumé and Singh, Aarti},
  volume = 	 {119},
  series = 	 {Proceedings of Machine Learning Research},
  month = 	 {13--18 Jul},
  publisher =    {PMLR},
  url = 	 {https://proceedings.mlr.press/v119/dutta20a.html},
}

@inproceedings{wick_unlocking_2019,
 author = {Wick, Michael and panda, swetasudha and Tristan, Jean-Baptiste},
 booktitle = {Advances in Neural Information Processing Systems},
 editor = {H. Wallach and H. Larochelle and A. Beygelzimer and F. d\textquotesingle Alch\'{e}-Buc and E. Fox and R. Garnett},
 pages = {},
 publisher = {Curran Associates, Inc.},
 title = {Unlocking Fairness: a Trade-off Revisited},
 url = {https://proceedings.neurips.cc/paper_files/paper/2019/file/373e4c5d8edfa8b74fd4b6791d0cf6dc-Paper.pdf},
 volume = {32},
 year = {2019}
}

@inproceedings{chen_why_2018,
 author = {Chen, Irene and Johansson, Fredrik D and Sontag, David},
 booktitle = {Advances in Neural Information Processing Systems},
 editor = {S. Bengio and H. Wallach and H. Larochelle and K. Grauman and N. Cesa-Bianchi and R. Garnett},
 pages = {},
 publisher = {Curran Associates, Inc.},
 title = {Why Is My Classifier Discriminatory?},
 url = {https://proceedings.neurips.cc/paper_files/paper/2018/file/1f1baa5b8edac74eb4eaa329f14a0361-Paper.pdf},
 volume = {31},
 year = {2018}
}

@misc{proptaxlocal,
title={Unpacking the State and Local Tax Toolkit: Sources of State and Local Tax Collections (FY 2020)}, 
author={Katherine Loughead and Jared Walczak and Eddie Koranyi},
year={2022}, 
month={08},
url={https://taxfoundation.org/data/all/state/state-local-tax-collections/}, 
howpublished={Tax Foundation}
}

@article{zhao_inherent_2022,
  author  = {Han Zhao and Geoffrey J. Gordon},
  title   = {Inherent Tradeoffs in Learning Fair Representations},
  journal = {Journal of Machine Learning Research},
  year    = {2022},
  volume  = {23},
  number  = {57},
  pages   = {1--26},
  url     = {http://jmlr.org/papers/v23/21-1427.html}
}

@misc{iaaostandards, 
title={STANDARD ON Mass Appraisal of Real Property: A criterion for measuring fairness 
quality, equity and accuracy}, 
year={2017}, 
url={https://www.iaao.org/wp-content/uploads/StandardOnMassAppraisal.pdf}, 
howpublished={International Association of Assessing Officers}}

@misc{compas, 
title={Machine Bias},
url={https://www.propublica.org/article/machine-bias-risk-assessments-in-criminal-sentencing}, 
authors={Julia Angwin and Jeff Larson and Surya Mattu and Lauren Kirchner},
year={2016}, 
month={05},
howpublished={ProPublica}
}

@article{harris2004assessing,
title={`Assessing' Discrimination: The Influence of Race in Residential Property Tax Assessments},
journal={Journal of Land Use \& Environmental Law},
volume={20},
number={1},
author={Lee Harris},
year={2004}, 
url={https://law.fsu.edu/sites/g/files/upcbnu1581/files/JLUEL/jluel-v20n1.pdf}}

@misc{nyc_report,
title={THE ROAD TO REFORM: A Blueprint for Modernizing and Simplifying New York City’s Property Tax System},
url={https://www.nyc.gov/assets/propertytaxreform/downloads/pdf/final-report.pdf}, 
author={Marc V. Shaw}, 
month={12},
year={2021}}

@misc{cook_county_treasurer_25,
title= {A Broken Property Tax Appeals System},
url={https://www.cookcountytreasurer.com/pdfs/appealsreportanalysis/AppealsReportAnalysisEnglish.pdf},
author={Cook County Treasurer's Office},
month={05},
year={2025}
}

@misc{cst_fbi, 
title={Former Cook County assessor official gets probation after helping feds},
url={https://chicago.suntimes.com/2025/07/29/former-cook-county-assessors-official-gets-probation-after-helping-with-feds-lengthy-probe}, 
month={07},
year={2025}, 
author={Jon Seidel},
howpublished={Chicago Sun-Times}}

@misc{abc_regressivity, 
title={`Highballed': How disproportionate property taxes are forcing some Americans out of their homes}, 
url={https://abcnews.go.com/US/highballed-disproportionate-property-taxes-forcing-americans-homes/story?id=124312846},
author={Mark Nichols and Tonya Simpson and Maia Rosenfeld and Jared Kofsky and Tommy Brooksbank},
howpublished={ABC News},
month={08},
year={2025},
}

@article{hayashi2014property,
title={Property Taxes and Their Limits: Evidence from New York City},
url={https://law.stanford.edu/publications/property-taxes-limits-evidence-new-york-city/},
journal={Stanford Law \& Policy Review},
volume={25},
author={Andrew T. Hayashi},
year={2014},
month={01}
}

@article{bidansetSurveyUseAutomated2022,
	title = {Survey on the use of automated valuation models ({AVMs}) in government assessment offices: {An} analysis of {AVM} use, acceptance, and barriers to more widespread implementation},
	volume = {19},
	issn = {3067-4816},
	shorttitle = {Survey on the use of automated valuation models ({AVMs}) in government assessment offices},
	url = {https://researchexchange.iaao.org/jptaa/vol19/iss2/3},
	doi = {10.63642/1357-1419.1250},
	language = {en},
	number = {2},
	urldate = {2025-08-15},
	journal = {Journal of Property Tax Assessment  \& Administration},
	author = {Bidanset, P. E. and Rakow, Ron},
	month = nov,
	year = {2022},
}

@article{dornfestStateProvincialProperty,
	title = {State and {Provincial} {Property} {Tax} {Policies} and {Administrative} {Practices} ({PTAPP}): 2017 {Findings} and {Report}},
	volume = {16},
	language = {en},
	number = {1},
	author = {Dornfest, Alan S and Rearich, Jennifer and Brydon III, T Douglas and Almy, Richard},
year={2019},
journal={Journal of Property Tax Assessment \& Administration},
url={https://doi.org/10.63642/1357-1419.1211}
}

@misc{ct_regressivity,
title={The Tax Divide | An unfair burden}, 
url={https://apps.chicagotribune.com/news/watchdog/cook-county-property-tax-divide/assessments.html},
author={Jason Grotto},
howpublished={Chicago Tribune},
month={06},
year={2017},
}

@misc{ct_regressivity2,
title={The Tax Divide | The problem with appeals}, 
url={https://apps.chicagotribune.com/news/watchdog/cook-county-property-tax-divide/appeals.html},
author={Jason Grotto},
howpublished={Chicago Tribune},
month={06},
year={2017},
}

@misc{ct_regressivity3,
title={The Tax Divide | An era of errors}, 
url={https://apps.chicagotribune.com/news/watchdog/cook-county-property-tax-divide/houlihan.html},
author={Jason Grotto},
howpublished={Chicago Tribune},
month={06},
year={2017},
}

@misc{ct_regressivity4,
title={The Tax Divide | Commercial breakdown}, 
url={https://apps.chicagotribune.com/news/watchdog/cook-county-property-tax-divide/index.html},
author={Jason Grotto and Sandhya Kambhampati},
howpublished={Chicago Tribune},
month={06},
year={2017},
}

@article{berryReassessingPropertyTax2021,
	title = {Reassessing the {Property} {Tax}},
	issn = {1556-5068},
	url = {https://www.ssrn.com/abstract=3800536},
	doi = {10.2139/ssrn.3800536},
	language = {en},
	urldate = {2025-06-02},
	journal = {SSRN Electronic Journal},
	author = {Berry, Christopher R.},
	month = mar,
	year = {2021},
}

@techreport{amornsiripanitchWhyAreResidential2022,
	type = {Working paper ({Federal} {Reserve} {Bank} of {Philadelphia})},
	title = {Why {Are} {Residential} {Property} {Tax} {Rates} {Regressive}?},
	url = {https://www.philadelphiafed.org/-/media/frbp/assets/working-papers/2022/wp22-02.pdf},
	number = {22-02},
	urldate = {2025-06-02},
	institution = {Federal Reserve Bank of Philadelphia},
	author = {Amornsiripanitch, Natee},
	month = jan,
	year = {2022},
	doi = {10.21799/frbp.wp.2022.02},
	pages = {22--02}
}

@article{avenancio-leonAssessmentGapRacial2022,
	title = {The {Assessment} {Gap}: {Racial} {Inequalities} in {Property} {Taxation}},
	volume = {137},
	copyright = {https://academic.oup.com/journals/pages/open\_access/funder\_policies/chorus/standard\_publication\_model},
	issn = {0033-5533, 1531-4650},
	shorttitle = {The {Assessment} {Gap}},
	url = {https://academic.oup.com/qje/article/137/3/1383/6522186},
	doi = {10.1093/qje/qjac009},
	language = {en},
	number = {3},
	urldate = {2025-06-02},
	journal = {The Quarterly Journal of Economics},
	author = {Avenancio-León, Carlos F and Howard, Troup},
	month = jul,
	year = {2022},
	pages = {1383--1434},
}

@techreport{perryDevaluationAssetsBlack2018,
	title = {The {Devaluation} of {Assets} in {Black} {Neighborhoods}},
	url = {https://www.brookings.edu/articles/devaluation-of-assets-in-black-neighborhoods/},
	language = {en},
	institution = {The Brookings Institution},
	author = {Perry, Andre and Rothwell, Jonathan and Harshbarger, David},
	month = nov,
	year = {2018},
}

@article{mcmillenMeasuresVerticalInequality2023,
	title = {Measures of vertical inequality in assessments},
	volume = {61},
	copyright = {https://www.elsevier.com/tdm/userlicense/1.0/},
	issn = {1051-1377},
	url = {https://linkinghub.elsevier.com/retrieve/pii/S1051137723000372},
	doi = {10.1016/j.jhe.2023.101950},
	language = {en},
	urldate = {2025-07-09},
	journal = {Journal of Housing Economics},
	author = {McMillen, Daniel and Singh, Ruchi},
	month = sep,
	year = {2023},
	pages = {101950},
}

@article{mcmillenAssessmentRegressivityProperty2020,
	title = {Assessment {Regressivity} and {Property} {Taxation}},
	volume = {60},
	copyright = {http://www.springer.com/tdm},
	issn = {0895-5638, 1573-045X},
	url = {http://link.springer.com/10.1007/s11146-019-09715-x},
	doi = {10.1007/s11146-019-09715-x},
	language = {en},
	number = {1-2},
	urldate = {2025-07-11},
	journal = {The Journal of Real Estate Finance and Economics},
	author = {McMillen, Daniel and Singh, Ruchi},
	month = feb,
	year = {2020},
	pages = {155--169},
}

@inproceedings{corbett-davies_algorithmic_2017,
	address = {New York, NY, USA},
	series = {{KDD}},
	title = {Algorithmic {Decision} {Making} and the {Cost} of {Fairness}},
	isbn = {978-1-4503-4887-4},
	url = {https://doi.org/10.1145/3097983.3098095},
	doi = {10.1145/3097983.3098095},
	urldate = {2023-04-06},
	booktitle = {Proceedings of the 23rd {ACM} {SIGKDD} {International} {Conference} on {Knowledge} {Discovery} and {Data} {Mining}},
	publisher = {Association for Computing Machinery},
	author = {Corbett-Davies, Sam and Pierson, Emma and Feller, Avi and Goel, Sharad and Huq, Aziz},
	month = aug,
	year = {2017},
	pages = {797--806},
}

@inproceedings{pleiss_fairness_2017,
	address = {Long Beach, CA},
	series = {{NIPS}},
	title = {On {Fairness} and {Calibration}},
	volume = {30},
	url = {https://papers.nips.cc/paper/2017/hash/b8b9c74ac526fffbeb2d39ab038d1cd7-Abstract.html},
	urldate = {2023-04-06},
	booktitle = {Advances in {Neural} {Information} {Processing} {Systems}},
	publisher = {Curran Associates, Inc.},
	author = {Pleiss, Geoff and Raghavan, Manish and Wu, Felix and Kleinberg, Jon and Weinberger, Kilian Q},
	month = dec,
	year = {2017},
}

@inproceedings{chouldechova_fair_2016,
	series = {{FATML}},
	title = {Fair prediction with disparate impact: {A} study of bias in recidivism prediction instruments},
	shorttitle = {Fair prediction with disparate impact},
	url = {http://arxiv.org/abs/1610.07524},
	doi = {10.48550/arXiv.1610.07524},
	urldate = {2023-04-06},
	booktitle = {Fairness, {Accountability}, and {Transparency} in {Machine} {Learning} ({FAT} {ML}) 2016},
	publisher = {arXiv},
	author = {Chouldechova, Alexandra},
	month = oct,
	year = {2016},
}

@inproceedings{kamiran_classifying_2009,
	series = {{IC4}},
	title = {Classifying without discriminating},
	doi = {10.1109/IC4.2009.4909197},
	booktitle = {2nd {International} {Conference} on {Computer} {Control} and {Communication} 2009},
	author = {Kamiran, Faisal and Calders, Toon},
	month = feb,
	year = {2009},
	pages = {1--6},
}

@inproceedings{hardt_equality_2016,
	series = {{NIPS}},
	title = {Equality of {Opportunity} in {Supervised} {Learning}},
	volume = {29},
	url = {https://papers.nips.cc/paper/2016/hash/9d2682367c3935defcb1f9e247a97c0d-Abstract.html},
	urldate = {2023-04-06},
	booktitle = {Advances in {Neural} {Information} {Processing} {Systems}},
	publisher = {Curran Associates, Inc.},
	author = {Hardt, Moritz and Price, Eric and Srebro, Nati},
	month = dec,
	year = {2016},
}

@inproceedings{zafar_fairness_2017,
	series = {{AISTATS}},
	title = {Fairness {Constraints}: {Mechanisms} for {Fair} {Classification}},
	shorttitle = {Fairness {Constraints}},
	url = {https://proceedings.mlr.press/v54/zafar17a.html},
	language = {en},
	urldate = {2023-04-06},
	booktitle = {Proceedings of the 20th {International} {Conference} on {Artificial} {Intelligence} and {Statistics}},
	publisher = {PMLR},
	author = {Zafar, Muhammad Bilal and Valera, Isabel and Rogriguez, Manuel Gomez and Gummadi, Krishna P.},
	month = apr,
	year = {2017},
	pages = {962--970},
}

@inproceedings{kleinberg_inherent_2017,
	series = {{ITCS}},
	title = {Inherent {Trade}-{Offs} in the {Fair} {Determination} of {Risk} {Scores}},
	url = {http://arxiv.org/abs/1609.05807},
	doi = {10.48550/arXiv.1609.05807},
	urldate = {2023-04-06},
	booktitle = {Innovations in {Theoretical} {Computer} {Science} ({ITCS}) 2017},
	publisher = {arXiv},
	author = {Kleinberg, Jon and Mullainathan, Sendhil and Raghavan, Manish},
	month = jan,
	year = {2017},
}

@inproceedings{black_algorithmic_2022,
	address = {Seoul Republic of Korea},
	series = {{FAccT}},
	title = {Algorithmic {Fairness} and {Vertical} {Equity}: {Income} {Fairness} with {IRS} {Tax} {Audit} {Models}},
	isbn = {978-1-4503-9352-2},
	shorttitle = {Algorithmic {Fairness} and {Vertical} {Equity}},
	url = {https://dl.acm.org/doi/10.1145/3531146.3533204},
	doi = {10.1145/3531146.3533204},
	language = {en},
	urldate = {2023-04-07},
	booktitle = {2022 {ACM} {Conference} on {Fairness}, {Accountability}, and {Transparency}},
	publisher = {ACM},
	author = {Black, Emily and Elzayn, Hadi and Chouldechova, Alexandra and Goldin, Jacob and Ho, Daniel},
	month = jun,
	year = {2022},
	pages = {1479--1503},
}

@inproceedings{chouldechova_case_2018,
	series = {{FAT}*},
	title = {A case study of algorithm-assisted decision making in child maltreatment hotline screening decisions},
	url = {https://proceedings.mlr.press/v81/chouldechova18a.html},
	language = {en},
	urldate = {2023-04-11},
	booktitle = {Proceedings of the 1st {Conference} on {Fairness}, {Accountability} and {Transparency}},
	publisher = {PMLR},
	author = {Chouldechova, Alexandra and Benavides-Prado, Diana and Fialko, Oleksandr and Vaithianathan, Rhema},
	month = jan,
	year = {2018},
	pages = {134--148},
}

@inproceedings{berk_convex_2017,
	address = {Halifax, Nova Scotia, Canada},
	series = {{FATML}},
	title = {A {Convex} {Framework} for {Fair} {Regression}},
	url = {http://arxiv.org/abs/1706.02409},
	doi = {10.48550/arXiv.1706.02409},
	urldate = {2023-04-11},
	booktitle = {{FAT} {ML}},
	publisher = {FAT ML},
	author = {Berk, Richard and Heidari, Hoda and Jabbari, Shahin and Joseph, Matthew and Kearns, Michael and Morgenstern, Jamie and Neel, Seth and Roth, Aaron},
	month = jun,
	year = {2017},
}

@inproceedings{feldman_certifying_2015,
	address = {New York, NY, USA},
	series = {{KDD}},
	title = {Certifying and {Removing} {Disparate} {Impact}},
	isbn = {978-1-4503-3664-2},
	url = {https://doi.org/10.1145/2783258.2783311},
	doi = {10.1145/2783258.2783311},
	urldate = {2023-04-11},
	booktitle = {Proceedings of the 21th {ACM} {SIGKDD} {International} {Conference} on {Knowledge} {Discovery} and {Data} {Mining}},
	publisher = {Association for Computing Machinery},
	author = {Feldman, Michael and Friedler, Sorelle A. and Moeller, John and Scheidegger, Carlos and Venkatasubramanian, Suresh},
	month = aug,
	year = {2015},
	pages = {259--268},
}

@article{kamiran_data_2012,
	title = {Data preprocessing techniques for classification without discrimination},
	volume = {33},
	issn = {0219-3116},
	url = {https://doi.org/10.1007/s10115-011-0463-8},
	doi = {10.1007/s10115-011-0463-8},
	language = {en},
	number = {1},
	urldate = {2023-04-11},
	journal = {Knowledge and Information Systems},
	author = {Kamiran, Faisal and Calders, Toon},
	month = oct,
	year = {2012},
	pages = {1--33},
}

@inproceedings{menon_cost_2018,
	series = {{FAT}*},
	title = {The cost of fairness in binary classification},
	url = {https://proceedings.mlr.press/v81/menon18a.html},
	language = {en},
	urldate = {2023-04-11},
	booktitle = {Proceedings of the 1st {Conference} on {Fairness}, {Accountability} and {Transparency}},
	publisher = {PMLR},
	author = {Menon, Aditya Krishna and Williamson, Robert C.},
	month = jan,
	year = {2018},
	pages = {107--118},
}

@inproceedings{quinonero_candela_disentangling_2023,
	address = {New York, NY, USA},
	series = {{FAccT}},
	title = {Disentangling and {Operationalizing} {AI} {Fairness} at {LinkedIn}},
	isbn = {9798400701924},
	url = {https://dl.acm.org/doi/10.1145/3593013.3594075},
	doi = {10.1145/3593013.3594075},
	urldate = {2023-06-13},
	booktitle = {Proceedings of the 2023 {ACM} {Conference} on {Fairness}, {Accountability}, and {Transparency}},
	publisher = {Association for Computing Machinery},
	author = {Quiñonero Candela, Joaquin and Wu, Yuwen and Hsu, Brian and Jain, Sakshi and Ramos, Jennifer and Adams, Jon and Hallman, Robert and Basu, Kinjal},
	month = jun,
	year = {2023},
	pages = {1213--1228},
}

@inproceedings{cooper_emergent_2021,
	address = {New York, NY, USA},
	series = {{AIES}},
	title = {Emergent {Unfairness} in {Algorithmic} {Fairness}-{Accuracy} {Trade}-{Off} {Research}},
	isbn = {978-1-4503-8473-5},
	url = {https://doi.org/10.1145/3461702.3462519},
	doi = {10.1145/3461702.3462519},
	urldate = {2023-06-15},
	booktitle = {Proceedings of the 2021 {AAAI}/{ACM} {Conference} on {AI}, {Ethics}, and {Society}},
	publisher = {Association for Computing Machinery},
	author = {Cooper, A. Feder and Abrams, Ellen},
	month = jul,
	year = {2021},
	pages = {46--54},
}

@article{koenecke_popular_2023,
	series = {{ICWSM}},
	title = {Popular {Support} for {Balancing} {Equity} and {Efficiency} in {Resource} {Allocation}: {A} {Case} {Study} in {Online} {Advertising} to {Increase} {Welfare} {Program} {Awareness}},
	volume = {17},
	issn = {2334-0770, 2162-3449},
	shorttitle = {Popular {Support} for {Balancing} {Equity} and {Efficiency} in {Resource} {Allocation}},
	url = {https://ojs.aaai.org/index.php/ICWSM/article/view/22163},
	doi = {10.1609/icwsm.v17i1.22163},
	language = {en},
	urldate = {2024-02-13},
	journal = {Proceedings of the International AAAI Conference on Web and Social Media},
	author = {Koenecke, Allison and Giannella, Eric and Willer, Robb and Goel, Sharad},
	month = jun,
	year = {2023},
	pages = {494--506},
}

@article{mice,
author = {Azur, Melissa J. and Stuart, Elizabeth A. and Frangakis, Constantine and Leaf, Philip J.},
title = {Multiple imputation by chained equations: what is it and how does it work?},
journal = {International Journal of Methods in Psychiatric Research},
volume = {20},
number = {1},
pages = {40-49},
doi = {https://doi.org/10.1002/mpr.329},
url = {https://onlinelibrary.wiley.com/doi/abs/10.1002/mpr.329},
eprint = {https://onlinelibrary.wiley.com/doi/pdf/10.1002/mpr.329},
year = {2011}
}

@misc{ccao_github_model,
year=2025, 
url={https://github.com/ccao-data/model-res-avm}, 
title={About
Automated valuation model for all class 200 residential properties in Cook County}, 
author={Cook County Assessor’s Office}}

@techreport{IAAO2018AVM,
  title        = {Standard on Automated Valuation Models (AVMs)},
  author       = {{International Association of Assessing Officers (IAAO)}},
  institution  = {International Association of Assessing Officers},
  address      = {Kansas City, MO},
  year         = {2018},
  month        = jul,
  type         = {Technical Standard (2nd edition)},
  url          = {https://www.iaao.org/wp-content/uploads/Standard_on_Automated_Valuation_Models.pdf}
}

@techreport{IAAO2021DataQuality,
  title        = {Standard on Data Quality},
  author       = {{International Association of Assessing Officers (IAAO)}},
  institution  = {International Association of Assessing Officers},
  address      = {Kansas City, MO},
  year         = {2021},
  type         = {Technical Standard},
  url          = {https://www.iaao.org/wp-content/uploads/Standard_on_Data_Quality.pdf}
}

@techreport{IAAO2013RatioStudies,
  title        = {Standard on Ratio Studies},
  author       = {{International Association of Assessing Officers (IAAO)}},
  institution  = {International Association of Assessing Officers},
  address      = {Kansas City, MO},
  year         = {2013},
  type         = {Technical Standard},
  url          = {https://www.iaao.org/wp-content/uploads/Standard_on_Ratio_Studies.pdf}
}

@article{suits1977measurement,
  title={Measurement of tax progressivity},
  author={Suits, Daniel B},
  journal={The American Economic Review},
  volume={67},
  number={4},
  pages={747--752},
  year={1977},
  publisher={JSTOR}
}

@article{dornfest2019state,
  title={State and Provincial property tax policies and administrative practices (PTAPP): 2017 findings and report},
  author={Dornfest, Alan S and Rearich, Jennifer and Brydon III, T Douglas and Almy, Richard},
  journal={Journal of Property Tax Assessment \& Administration},
  volume={16},
  number={1},
  pages={43--130},
  year={2019}
}

@article{kokinis2005use,
  title={Use of the cost, income and sales-comparison approaches in the valuation of real estate},
  author={Kokinis-Graves, Carol},
  journal={J. St. Tax'n},
  volume={24},
  pages={23},
  year={2005},
  publisher={HeinOnline}
}

@article{jennifer2021garbage,
  title={Garbage in, garbage out: Implications of data quality for valuation models},
  author={Jennifer, MAS and RES, Rearich and others},
  journal={Journal of Property Tax Assessment \& Administration},
  volume={18},
  number={1},
  pages={1},
  year={2021}
}

@article{bidanset2017accounting,
  title={Accounting for locational, temporal, and physical similarity of residential sales in mass appraisal modeling: the development and application of geographically, temporally, and characteristically weighted regression},
  author={Bidanset, Paul E and McCord, Michael and Lombard, John R and Davis, Peadar and McCluskey, William},
  journal={Journal of Property Tax Assessment \& Administration},
  volume={14},
  number={2},
  pages={5--13},
  year={2017}
}

@article{quintos2014improving,
  title={Improving assessment equity in mass appraisal models},
  author={Quintos, Carmela},
  journal={Journal of Property Tax Assessment \& Administration},
  volume={11},
  number={4},
  pages={53--64},
  year={2014}
}

@techreport{berry2025evaluation,
  title        = {An Evaluation of Progress on Residential 
Assessment Fairness in Cook County},
  author       = {Berry, Christopher},
  institution  = {Mansueto Institute for Urban Innovation, University of Chicago},
  address      = {Chicago, IL},
  year         = {2025},
  url          = {https://bpb-us-w2.wpmucdn.com/voices.uchicago.edu/dist/6/2330/files/2025/09/Kaegi-Evaluation-9_04.pdf}
}

@techreport{IAAO2018GISTechSurvey,
  title        = {Summary Findings of IAAO’s GIS Technology Survey for Assessment Professionals},
  institution  = {International Association of Assessing Officers},
  address      = {Kansas City, MO},
  year         = {2018},
  type         = {Technical Report},
  url          = {https://www.iaao.org/wp-content/uploads/GIS_SURVEY2018.pdf}
}

@article{droj2024comprehensive,
  title={A comprehensive overview regarding the impact of GIS on property valuation},
  author={Droj, Gabriela and Kwartnik-Pruc, Anita and Droj, Laurențiu},
  journal={ISPRS International Journal of Geo-Information},
  volume={13},
  number={6},
  pages={175},
  year={2024},
  publisher={MDPI}
}

@article{drometer2018wealth,
  title={Wealth and inheritance taxation: An overview and country comparison},
  author={Drometer, Marcus and Frank, Marco and P{\'e}rez, Maria Hofbauer and Rhode, Carla and Schworm, Sebastian and Stitteneder, Tanja},
  journal={ifo DICE Report},
  volume={16},
  number={2},
  pages={45--54},
  year={2018},
  publisher={M{\"u}nchen: ifo Institut-Leibniz-Institut f{\"u}r Wirtschaftsforschung an der~…}
}

@article{ho2020affirmative,
  title={Affirmative algorithms: The legal grounds for fairness as awareness},
  author={Ho, Daniel E and Xiang, Alice},
  journal={U. Chi. L. Rev. Online},
  pages={134},
  year={2020},
  publisher={HeinOnline}
}

@article{harvey2025cotality,
    title={Hidden Errors in Big Data: Brokered Data Obfuscates Measures of Vertical Equity},
    author={Harvey, Emma and Smith, Evelyn},
    Journal={Working paper},
    year={2025}
}

@article{ihlanfeldt2022homestead,
  title={Homestead exemptions, heterogeneous assessment, and property tax progressivity},
  author={Ihlanfeldt, Keith and Rodgers, Luke P},
  journal={National Tax Journal},
  volume={75},
  number={1},
  pages={7--31},
  year={2022},
  publisher={The University of Chicago Press}
}

@article{oates2016local,
  title={Are local property taxes regressive, progressive, or what?},
  author={Oates, Wallace E and Fischel, William A},
  journal={National Tax Journal},
  volume={69},
  number={2},
  pages={415--433},
  year={2016},
  publisher={The University of Chicago Press}
}

@article{fan2006determinants,
  title={Determinants of house price: A decision tree approach},
  author={Fan, Gang-Zhi and Ong, Seow Eng and Koh, Hian Chye},
  journal={Urban Studies},
  volume={43},
  number={12},
  pages={2301--2315},
  year={2006},
  publisher={Sage Publications Sage UK: London, England}
}

@article{yilmazer2020mass,
  title={A mass appraisal assessment study using machine learning based on multiple regression and random forest},
  author={Yilmazer, Seckin and Kocaman, Sultan},
  journal={Land use policy},
  volume={99},
  pages={104889},
  year={2020},
  publisher={Elsevier}
}

@article{hong2020house,
  title={A house price valuation based on the random forest approach: the mass appraisal of residential property in South Korea},
  author={Hong, Jengei and Choi, Heeyoul and Kim, Woo-sung},
  journal={International Journal of Strategic Property Management},
  volume={24},
  number={3},
  pages={140--152},
  year={2020}
}

@article{hanlon2005book,
  title={Book-tax conformity for corporate income: An introduction to the issues},
  author={Hanlon, Michelle and Shevlin, Terry},
  journal={Tax policy and the economy},
  volume={19},
  pages={101--134},
  year={2005},
  publisher={The MIT Press}
}

@incollection{kahrl2024black,
  title={Citizens and taxpayers},
  author={Kahrl, Andrew W},
  booktitle={The Black tax: 150 years of theft, exploitation, and dispossession in America},
  year={2024},
  publisher={University of Chicago Press}
}

@misc{wasi2005property,
  title={Property tax limitations and mobility: the lock-in effect of California's Proposition 13},
  author={Wasi, Nada and White, Michelle J},
  year={2005},
  publisher={National Bureau of Economic Research Cambridge, Mass., USA}
}

@article{dornfest2014state,
  title={State and provincial property tax policies and administrative practices (PTAPP): 2012 update of 2009 compilation and report},
  author={Dornfest, Alan S and Van Sant, Steve and Anderson, Rick},
  journal={Journal of Property Tax Assessment \& Administration},
  volume={11},
  number={3},
  pages={15--116},
  year={2014}
}

@INPROCEEDINGS{elzayn_estimating_2024,
  author={Elzayn, Hadi and Black, Emily and Vossler, Patrick and Jo, Nathanael and Goldin, Jacob and Ho, Daniel E.},
  booktitle={2024 IEEE Conference on Secure and Trustworthy Machine Learning (SaTML)}, 
  title={Estimating and Implementing Conventional Fairness Metrics With Probabilistic Protected Features}, 
  year={2024},
  volume={},
  number={},
  pages={161-193},
  doi={10.1109/SaTML59370.2024.00016}}

@article{Elzayn_measuring_2024,
    author = {Elzayn, Hadi and Smith, Evelyn and Hertz, Thomas and Guage, Cameron and Ramesh, Arun and Fisher, Robin and Ho, Daniel E and Goldin, Jacob},
    title = {Measuring and Mitigating Racial Disparities in Tax Audits*},
    journal = {The Quarterly Journal of Economics},
    volume = {140},
    number = {1},
    pages = {113-163},
    year = {2024},
    month = {09},
    issn = {0033-5533},
    doi = {10.1093/qje/qjae027},
    url = {https://doi.org/10.1093/qje/qjae027},
    eprint = {https://academic.oup.com/qje/article-pdf/140/1/113/60122543/qjae027.pdf},
}

@article{berk2017convex,
  title={A convex framework for fair regression},
  author={Berk, Richard and Heidari, Hoda and Jabbari, Shahin and Joseph, Matthew and Kearns, Michael and Morgenstern, Jamie and Neel, Seth and Roth, Aaron},
  journal={arXiv preprint arXiv:1706.02409},
  year={2017}
}

@article{beebe2025texas,
  title={The Surprising Impact of the 20\% Appraisal Cap in Texas},
  author={Beebe, Joyce and Diamond, John W. and Rabb, Jennifer},
  journal={Center for Tax and Budget Policy Research Paper},
  publisher={Baker Institute for Public Policy},
  year={2025},
  url={https://www.bakerinstitute.org/research/surprising-impact-20-appraisal-cap-texas}
}

@article{mcmillen2020assessment,
  title={Assessment regressivity and property taxation},
  author={McMillen, Daniel and Singh, Ruchi},
  journal={The Journal of Real Estate Finance and Economics},
  volume={60},
  number={1},
  pages={155--169},
  year={2020},
  publisher={Springer}
}

@article{edelstein1979appraisal,
  title={An appraisal of residential property tax regressivity},
  author={Edelstein, Robert H},
  journal={Journal of Financial and Quantitative Analysis},
  volume={14},
  number={4},
  pages={753--768},
  year={1979},
  publisher={Cambridge University Press}
}

@article{hodge2017assessment,
  title={Assessment inequity in a declining housing market: The case of Detroit},
  author={Hodge, Timothy R and McMillen, Daniel P and Sands, Gary and Skidmore, Mark},
  journal={Real Estate Economics},
  volume={45},
  number={2},
  pages={237--258},
  year={2017},
  publisher={Wiley Online Library}
}

@article{lin2010property,
  title={Property tax inequity resulting from inaccurate assessment—The Taiwan experience},
  author={Lin, Tzu-Chin},
  journal={Land Use Policy},
  volume={27},
  number={2},
  pages={511--517},
  year={2010},
  publisher={Elsevier}
}

@article{geraci1977measuring,
  title={Measuring the benefits from property tax assessment reform},
  author={Geraci, Vincent J},
  journal={National Tax Journal},
  volume={30},
  number={2},
  pages={195--205},
  year={1977},
  publisher={The University of Chicago Press}
}

@article{hou2025assessment,
  title={Assessment frequency and equity of the property tax: Latest evidence from Philadelphia},
  author={Hou, Yilin and Ding, Lei and Schwegman, David J and Barca, Alaina G},
  journal={Journal of Policy Analysis and Management},
  volume={44},
  number={2},
  pages={483--507},
  year={2025},
  publisher={Wiley Online Library}
}

@article{weber2010ask,
  title={Ask and ye shall receive? Predicting the successful appeal of property tax assessments},
  author={Weber, Rachel N and McMillen, Daniel P},
  journal={Public Finance Review},
  volume={38},
  number={1},
  pages={74--101},
  year={2010},
  publisher={SAGE Publications Sage CA: Los Angeles, CA}
}

@article{mcmillen2013effect,
  title={The effect of appeals on assessment ratio distributions: Some nonparametric approaches},
  author={McMillen, Daniel P},
  journal={Real Estate Economics},
  volume={41},
  number={1},
  pages={165--191},
  year={2013},
  publisher={Wiley Online Library}
}

@article{bayer2017racial,
  title={Racial and ethnic price differentials in the housing market},
  author={Bayer, Patrick and Casey, Marcus and Ferreira, Fernando and McMillan, Robert},
  journal={Journal of Urban Economics},
  volume={102},
  pages={91--105},
  year={2017},
  publisher={Elsevier}
}

@article{wagstaff2012machine,
  title={Machine learning that matters},
  author={Wagstaff, Kiri},
  journal={arXiv preprint arXiv:1206.4656},
  year={2012}
}

@inproceedings{black2023toward,
  title={Toward operationalizing pipeline-aware ML fairness: A research agenda for developing practical guidelines and tools},
  author={Black, Emily and Naidu, Rakshit and Ghani, Rayid and Rodolfa, Kit and Ho, Daniel and Heidari, Hoda},
  booktitle={Proceedings of the 3rd ACM Conference on Equity and Access in Algorithms, Mechanisms, and Optimization},
  pages={1--11},
  year={2023}
}
\clearpage

\appendix

\section{Regressivity Metrics \label{app:reg_metrics}}

\subsection{Suits Index}

The Suits Index, introduced by \cite{suits1977measurement} as a tool for evaluating tax progressivity, is closely related to the Gini coefficient in its reliance on Lorenz-style curves. Unlike the Gini, however, it is derived from a single cumulative curve. The steps for computing the index are:

\begin{enumerate} 
    \item Rank homes in ascending order by sale price.
    \item For this ordered list, compute cumulative shares of both sale prices and assessed values, ranging from 0 to 100 percent.
    \item Plot the cumulative share of assessed values (y-axis) against the cumulative share of sale prices (x-axis).
    \item Define the index as      \begin{equation} 
        S = \frac{(K-L)}{K} \end{equation} 
    
   which represents the area beneath the 45-degree line less the area under the cumulative assessment curve. Because both axes span 0–100, $K = 0.5*100*100 = 5,000$. Thus, the Suits Index simplifies to $S = 1-\frac{1}{5000}*L$. 
\end{enumerate}

We approximate $L$ using the discrete method outlined in \cite{mcmillenMeasuresVerticalInequality2023}. An index value of 0 reflects perfectly proportional assessments ($K=L$). Negative values arise when $L>K$, meaning that, for example, the bottom 10\% of homes by sale price represent more than 10\% of total assessed value, which is taken as evidence of regressivity. Positive values, in contrast, indicate progressivity.

\subsection{Price-Related Differential (PRD)}

The price-related differential, or PRD, is a measure which compares the average assessment ratio to the sale price weighted average ratio among the population of asssessed homes. Formally, the PRD is given by:

\begin{equation}
    PRD = \frac{\bar{R}}{\sum_{i}(\frac{{A_i}}{{S_i}})(\frac{S_i}{\sum_j{S_j}})} = \frac{\bar{R}}{\bar{A}/\bar{S}}
\end{equation}

Where $\bar{R}$ is the average assessment ratio, $\bar{A}$ the average assessed value, and $\bar{S}$ the average sale price. PRD values greater than 1 indicate that high-priced homes have lower assessment ratios than low-priced homes, which is an indication of regressivity. Per IAAO standards, the acceptable range for the PRD is between 0.98 and 1.03 \cite{IAAO2013RatioStudies}.

\subsection{Regression-Based Measures}

Regression-based measures of assessment regressivity capture linear associations between sale price and assessed value. Numerous specifications have been adopted in the assessment literature, including regressions of sale price on assessed value, of assessed value on sale price, and of sales ratios on sale price, alongside logged versions of these specifications and more complicated formulations such as the coefficient of price-related bias \citep{IAAO2013RatioStudies}.

The measure used in the proof provided in Appendix \ref{app:tradeoff} is derived from a regression of log sales ratios on log sale prices. More formally, the measure is estimated as the coefficient $\beta_1$ from the following expression:

\begin{equation}
    log(A/S) = \beta_0 + \beta_1 log(S) + \varepsilon
\end{equation}

Where $A$ is assessed value and $S$ is sale price. Intuitively, if assessments are regressive, then sales ratios will generally decrease as sale prices increase, and $\beta_1$ will be negative.

\cite{mcmillenMeasuresVerticalInequality2023} uses simulated data to establish that regression-based measures of assessment inequality are generally biased downwards, towards finding regressivity. This bias stems from a number of sources, including measurement error in sale price, endogeneity bias, and the functional form of assessment models. However, \cite{mcmillenMeasuresVerticalInequality2023} also establish that regression-based measures are sensitive to changes in simulated regressivity even if the levels are biased. The majority of the applications of this metric in this paper, in particular in \S\ref{s-4-results} and Appendix \ref{app:tradeoff}, consider changes in regressivity between assessment models rather than the overall level of regressivity. If the extent of bias is uncorrelated with the method of assessment, as is the case for measurement error in sale price, then the bias will difference out, and the resulting estimate of the change in regressivity will be valid. In cases where we examine the level of regressivity, we show robustness of these results to other metrics.
\newpage
\section{Assessment Accuracy and Effective Tax Rates \label{app:accuracy_etr}}

In this section, we illustrate the relationship between our primary assessment accuracy metric (MAPE) and property tax rates. Specifically, we show that the percentage difference between a property's \textit{statutory} and \textit{effective} tax rates is equivalent to the mean absolute percentage error between assessed value and sale price in a setting with no exemptions or deductions to the property tax.

First, let the statutory property tax rate\textemdash the nominal rate of tax established by state or local law\textemdash be denoted $\tau$. Similarly, let $\hat{\tau}$ denote the effective tax rate\textemdash the actual rate of tax paid relative to a property's underlying market value. Let $A$ denote assessed value, and $S$ denote market value. If there are no exemptions or deductions to the tax, we have that:

\begin{equation}
    \tau{A} = \hat{\tau}S
\end{equation}

That is, the statutory rate applied to the assessed value of the home is equal to the effective tax rate applied to the home's underlying market value.

Re-arranging and subtracting 1 from both sides yields:

\begin{equation}
    \frac{A}{S} = \frac{\hat{\tau}}{\tau} \iff \frac{A}{S} -1 = \frac{\hat{\tau}}{\tau} -1 \iff \frac{A-S}{S} = \frac{\hat{\tau} - \tau}{\tau}
\end{equation}

The left-hand side of the final term is the percentage error between assessed value and sale price, while the right-hand side is the percentage error between effective and statutory property tax rates. One can therefore evaluate mean absolute percentage error across the population using either term and arrive at the same result.
\newpage
\section{Relationship Between Fairness and Accuracy: Log Coefficient and Mean Squared Error \label{app:tradeoff}}

This section provides a theoretical analysis the relationship between fairness and accuracy using the regression-based definition of model fairness described in Appendix \ref{app:reg_metrics} and mean squared error as the measure of model accuracy.

Consider two assessment models, denoted 1 and 2. Each model predicts ground-truth price, denoted $S$. The predictions of each model are given by $\hat{S}_i$, $i \in \{1, 2\}$.

Let fairness be measured using the coefficient $\beta_1$ from the following regression:

\begin{equation}
    log(\frac{\hat{S_i}}{S}) = \beta_0 + \beta_1log(S) + \varepsilon
\end{equation}

If $\beta_1$ is negative, then the log ratio of assessed to actual sale price is decreasing in log sale price, indicating that assessments are regressive. Therefore, increases in $\beta_1$ signal that fairness is improving between one assessment model and the next. 

The formula for $\hat{\beta_1}$ is given by:

\begin{equation}
    \hat{\beta_1} = \frac{cov(log(\hat{S_i}/S), log(S))}{var(log(S))} = \frac{cov(log(\hat{S_i}), log(S))}{var(log(S))} -1
\end{equation}

Because $var(log(S))$, the variance of log ground-truth price, is unchanging between models, the key determinant of model fairness is the term in the numerator, $cov(log(\hat{S_i}), log(S))$. More specifically, if:

\begin{equation}
\label{eq:cov_ineq}
     cov(log(\hat{S}_2), log(S)) - cov(log(\hat{S}_1), log(S)) > 0
\end{equation}

Then the value of $\beta_1$ will be higher under model 2 than under model 1, and consequently model 2 will be fairer according to the regression coefficient metric than model 1.

Let $s$ denote $log(S)$ and $\hat{s}_i$ denote $log(\hat{S}_i)$. We can express the covariance between predicted and actual log prices for model $i$ as:

\begin{equation}
\label{eq:cov_expansion}
cov(\hat{s}_i, s) = \frac{1}{2}(\sigma^2_{\hat{s}_i} + (\mu_{\hat{s}_i}-\mu_{s})^2 - MSE(\hat{s}_i, s) + \sigma^2_s)
\end{equation}

To see this, note that the mean squared error between $\hat{s}_i$ and $s$ is given by:

\begin{multline}
    MSE(\hat{s}_i, s) = E[\hat{s}_i^2] - 2E[\hat{s}_is] + E[s^2] \\
    = \sigma^2_{\hat{s}_i} + \mu_{\hat{s}_i}^2 - 2(cov(\hat{s}_i,s) + \mu_{\hat{s}_i}\mu_s) + \sigma^2_s + \mu_s^2 \\
    = \sigma^2_{\hat{s}_i} + (\mu_{\hat{s}_i} - \mu_s)^2 -2cov(\hat{s}_i,s) + \sigma_s^2
\end{multline}

Which yields the expression in Equation \ref{eq:cov_expansion} after some re-arranging. Given this, we can express the difference in the covariance between predicted and actual values between model 2 and model 1 as:

\begin{equation}
    cov(\hat{s}_2, s) - cov(\hat{s}_1, s)  = \frac{1}{2} \left[ ( \sigma^2_{\hat{s}_2} -  \sigma^2_{\hat{s}_1}) + [(\mu_{\hat{s}_1} - \mu_s)^2 - (\mu_{\hat{s}_2} - \mu_s)^2] - [ MSE(\hat{s}_2, s) -  MSE(\hat{s}_1, s)] \right]
\end{equation}

It follows that the relationship between fairness and accuracy under these metrics depends on the relative size of the following three terms, which together determine the change in model fairness $cov(\hat{s}_2, s) - cov(\hat{s}_1, s)$:

\begin{enumerate}
    \item The change in the variance of model predictions, $( \sigma^2_{\hat{s}_2} -  \sigma^2_{\hat{s}_1})$,
    \item The change in the squared difference between mean predicted and mean ground-truth log price, $(\mu_{\hat{s}_1} - \mu_s)^2 - (\mu_{\hat{s}_2} - \mu_s)^2$, 
    \item And the change in model accuracy, $-[MSE(\hat{s}_2, s) -  MSE(\hat{s}_1, s)]$
\end{enumerate}

The relationship between fairness and accuracy in this setting is therefore ambiguous. To see this, assume that models 1 and 2 have the same mean prediction, $\mu_{\hat{s}_1} = \mu_{\hat{s}_2}=\mu_{\hat{s}}$. This sets the value of term (2) to zero. Further, let model 2 have lower mean squared error than model 1, so term (3) is positive. Whether fairness increases alongside accuracy now depends entirely on the difference in the variance of the predictions between models 1 and 2, given by term (1). If model 2's predictions are significantly less variable than model 1, and the magnitude of the difference exceeds the magnitude of model 2's accuracy gains, then model 2 will be less fair than model 1, despite being more accurate. If, however, model 2's predictions are more variable than model 1's predictions, or exhibit only slight decreases in variance relative to model 1, then accuracy and fairness will improve simultaneously.

In summary, simultaneous improvements in fairness and accuracy in the assessment setting are possible, as are tradeoffs, and the realized outcome depends on changes in the distribution of predictions relative to ground-truth prices.
\newpage
\section{Additional Tables and Figures \label{app:robustness}}

\begin{figure}[!htb]
    \centering

    \begin{subfigure}{0.45\textwidth}
        \includegraphics[width=\linewidth]{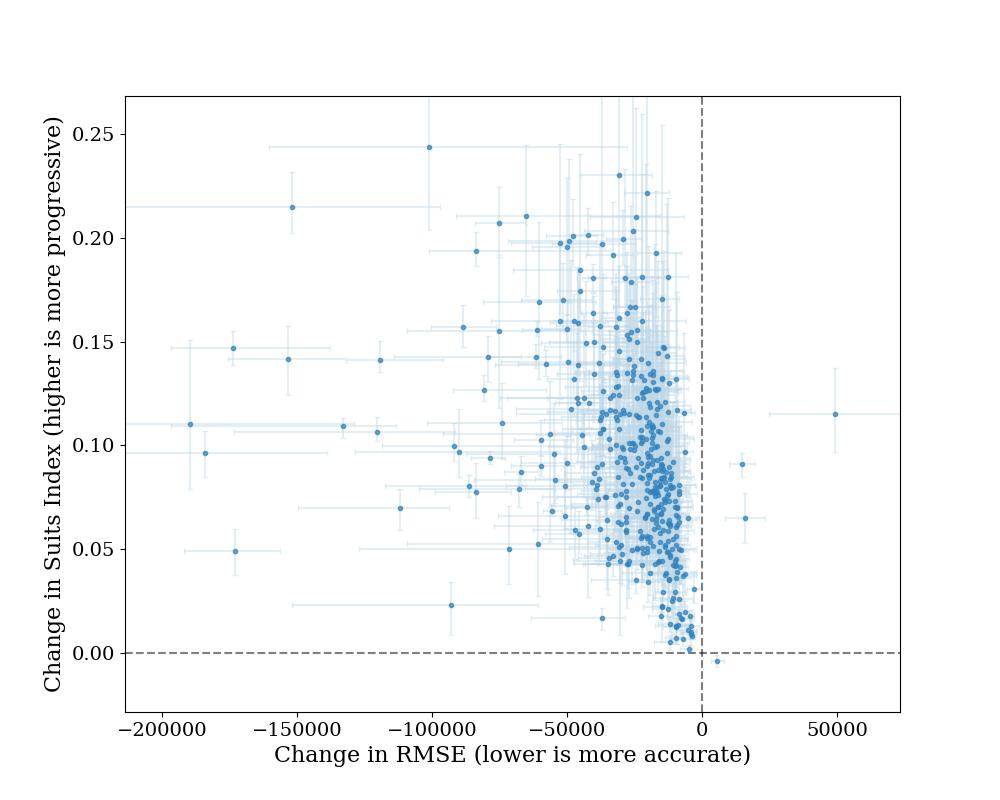}
        \label{fig:ablation_plots_alt_metrics_rmse_suits}
        \caption{RMSE, Suits Index}
    \end{subfigure}
 \hfill
    \begin{subfigure}{0.45\textwidth}
        \includegraphics[width=\linewidth]{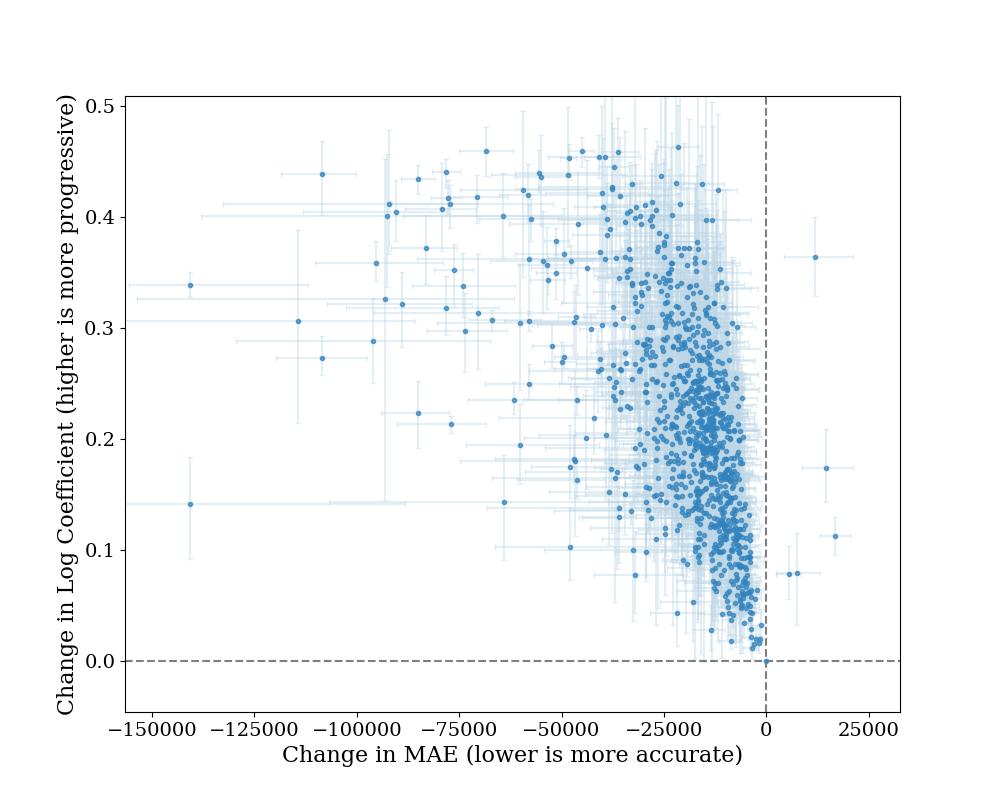}
        \label{fig:ablation_plots_alt_metrics_mae_coef}
        \caption{MAE, LC}
    \end{subfigure}

    \caption{Pareto Improvements in Model Performance after Adding More Property Characteristics to Assessment Models are Robust Across Metrics}
    \label{fig:ablation_plots_alt_metrics}

    \begin{tablenotes}[flushleft]
    
    \scriptsize 
    \item \textit{Notes:} The figures show changes in accuracy and fairness between the ``sparse" and ``rich" county-level LASSO-based assessment models described in Section \ref{s-4-results}. Models are trained using property and time-of-sale characteristics. Accuracy is measured using root mean squared error (left panel) and mean absolute error (right panel) between assessed and sale price. Fairness is measured using the Suits Index (left panel) and LC (right panel) measures described in \ref{app:reg_metrics}. Changes in performance are statistically significant at a 95\% confidence level after applying a Benjamini-Hochberg multiple testing correction. Confidence intervals are computed using a studentized bootstrap and displayed in light blue. Results are censored at the 2nd and 98th percentiles.
    \end{tablenotes}
    \Description{The figures show changes in accuracy and fairness for county-level assessment models across the ablation experiment. Both panels shows the changes between ``sparse" and ``rich" county-level LASSO-based assessment models. In both panels, accuracy is measured using MAPE between assessed and sale price. In the left panel, fairness is measured using the Suits Index; in the right panel, it is measured using Log Coefficient. Both figures show a trend of simultaneously increasing fairness and accuracy.}
\end{figure}

\begin{figure}[!htb]
    \centering

    \begin{subfigure}{0.45\textwidth}
        \includegraphics[width=\linewidth]{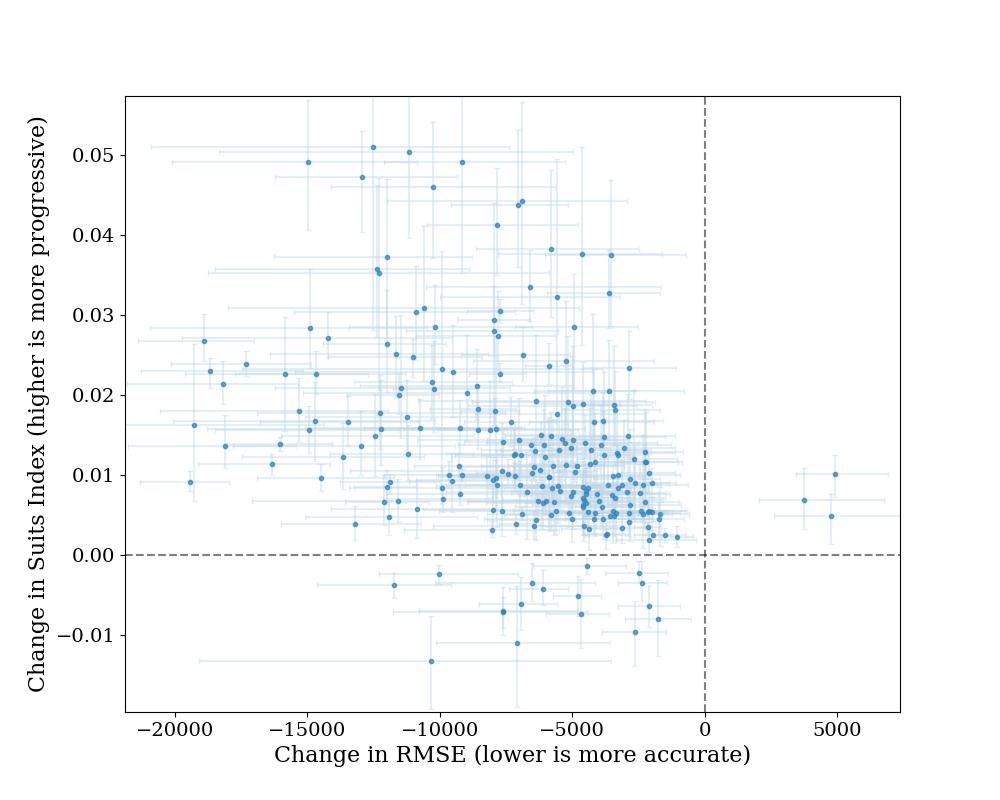}
        \label{fig:census_delta_alt_rmse_suits}
        \caption{RMSE, Suits Index}
    \end{subfigure}
 \hfill
    \begin{subfigure}{0.45\textwidth}
        \includegraphics[width=\linewidth]{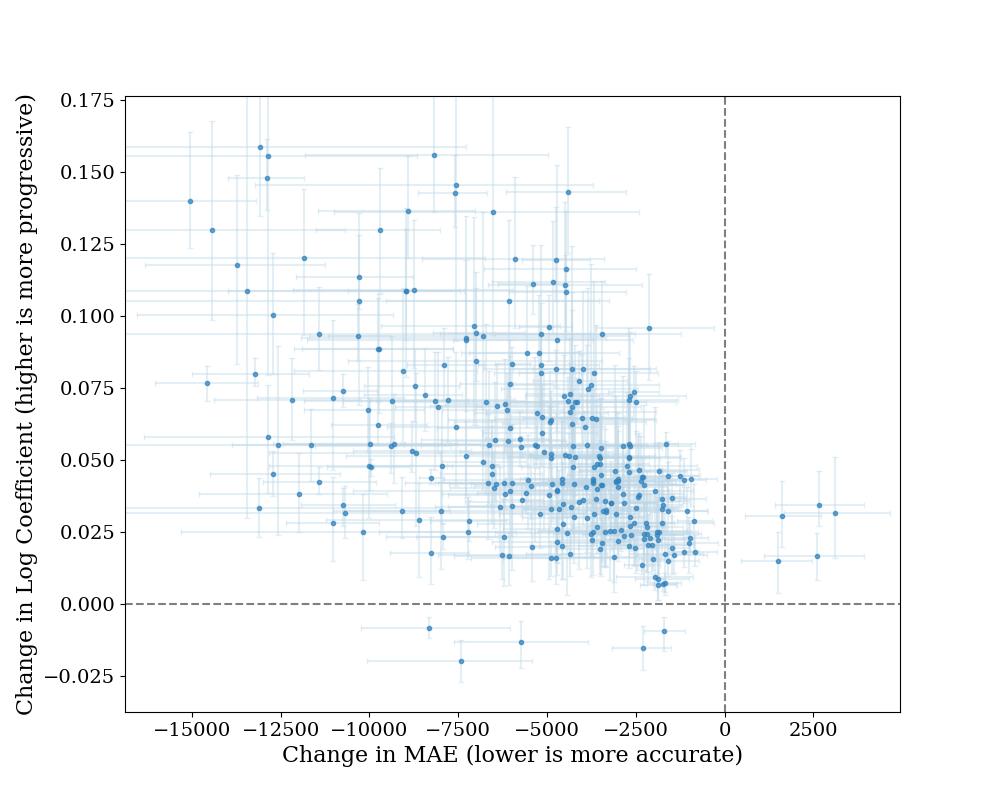}
        \label{fig:census_delta_alt_mae_lc}
        \caption{MAE, LC}
    \end{subfigure}

    \caption{Pareto Improvements in Model Performance after Incorporating Census Data are Robust Across Metrics}
    \label{fig:census_delta_alt}

    \begin{tablenotes}[flushleft]
    
    \scriptsize 
    \item \textit{Notes:} The figures show changes in accuracy and fairness for county-level random forest assessment models following the addition of Census characteristics. The baseline model for each county uses only status quo predictions and time-of-sale data. The alternative model incorporates these features in addition to Census block group characteristics. Accuracy is measured using the change in root mean squared error (left panel) and mean absolute error (right panel) between predicted and actual sale price as we move from the baseline model to the alternative model. Fairness is measured using the Suits Index and LC measures described in Appendix \ref{app:reg_metrics}. Changes in performance are statistically significant at a 95\% confidence level after applying a Benjamini-Hochberg multiple testing correction. Confidence intervals are computed using a studentized bootstrap and displayed in light blue. Results are censored at the 2nd and 98th percentiles.
    \end{tablenotes}
    \Description{The figures show changes in accuracy and fairness for county-level assessment models across the census experiment. Both panel shows  changes in accuracy and fairness after the addition of Census block group data to county-level random forest-based assessment models. In both panels, accuracy is measured using MAPE between assessed and sale price. In the left panel, fairness is measured using the Suits Index; in the right panel, it is measured using Log Coefficient. Both figures show a trend of simultaneously increasing fairness and accuracy.}
\end{figure}

\begin{figure}
    \centering

    \begin{subfigure}{0.8\textwidth}
        \includegraphics[width=\linewidth]{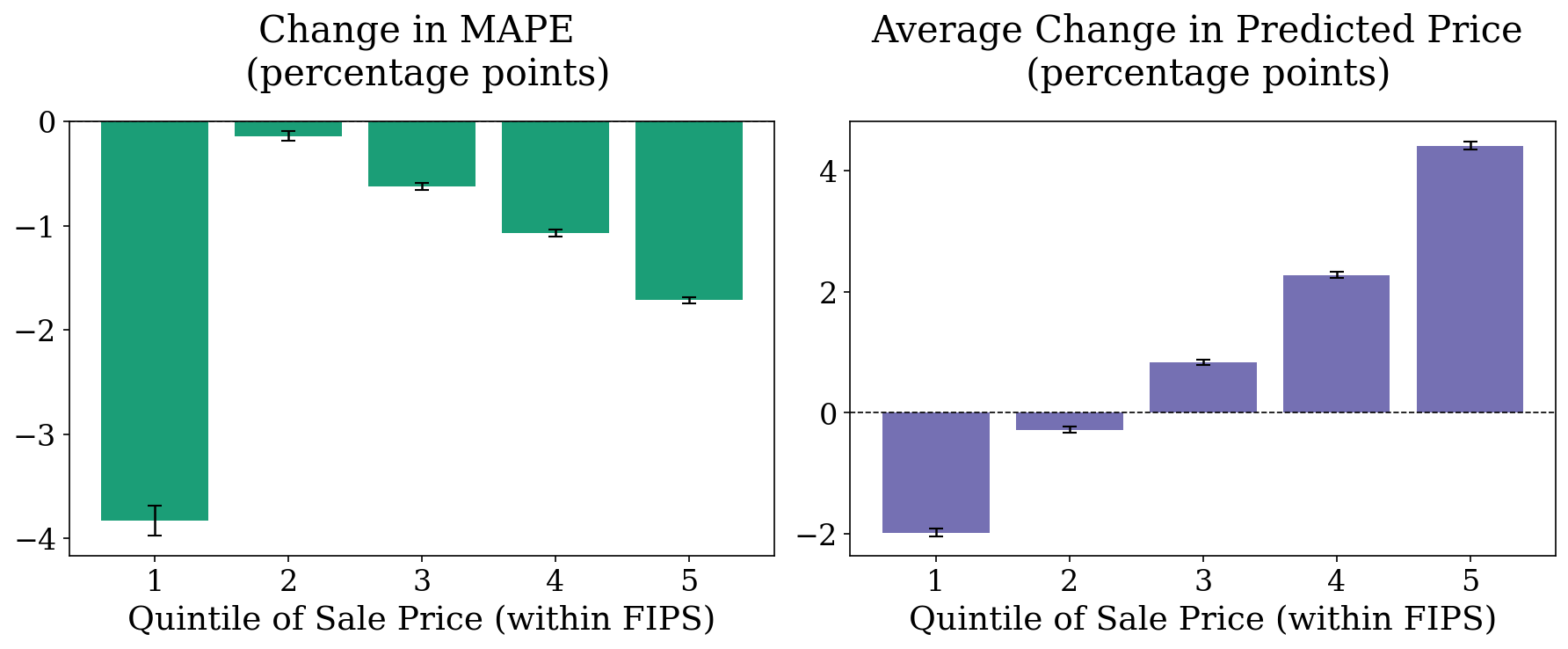}
        \label{}
    \end{subfigure}

    \caption{Impact of Census Characteristics on Assessment Accuracy and Levels, by Quintile of Sale Price}
    \label{fig:quintile_comparison}

    \begin{tablenotes}[flushleft]
    
    \scriptsize 
    \item \textit{Notes:} The figures shows the impact of Census characteristics on assessment accuracy and levels across all 1,619 counties in states without assessment caps. Changes are reported within quintiles of sale price, which are defined at the FIPS level. The left panel shows the average change in mean absolute percentage error (MAPE) and the right panel shows the average change in predicted price as we move from the baseline model to the Census characteristic model described in Section \ref{s-4.2-improvements}. Error bars represent 95\% confidence intervals computed using a percentile bootstrap.
    \end{tablenotes}
\end{figure}

\begin{table}
\centering

\begin{tabular}{lrr}
\hline
 & \textbf{Sold} & \textbf{Assessed} \\
\hline
N & 3,066,716 & 100,969,072 \\
Year built & 1,978.67 & 1,974.94 \\
 & (33.11) & (31.64) \\
Building size (square feet) & 1,885.05 & 1,876.19 \\
 & (2,144.16) & (2,173.59) \\
Lot size (square feet) & 1,425.76 & 1,417.52 \\
 & (1,061.11) & (2,185.51) \\
Number of stories & 1.37 & 1.35 \\
 & (0.70) & (0.70) \\
Number of bedrooms & 3.21 & 3.21 \\
 & (1.33) & (1.68) \\
Build quality & 6.43 & 6.34 \\
 & (1.98) & (2.04) \\
\hline
\end{tabular}
\vspace{4pt}
\caption{Characteristics of sold versus assessed properties, 2023}\label{tab:sold_assessed_summ_stats}

\begin{tablenotes}[flushleft]
    
    \scriptsize 
    \item \textit{Notes:} The table compares mean property characteristics for sold and assessed homes for 2023. Standard deviations are displayed in parentheses. The underlying data is pooled across all counties in Cotality's data. "Build quality" is an ordinal measure ranging from 1 to 9; higher values correspond to higher quality buildings.
\end{tablenotes}
\end{table}

\begin{table}[ht]
\centering

\small
\begin{tabular}{llcccc}
\toprule
\textbf{FIPS Code} & \textbf{County Name} &
\textbf{Fairness (Weighted)} & \textbf{Fairness (Unweighted)} &
\textbf{Error (Weighted)} & \textbf{Error (Unweighted)} \\
\midrule
8077 & Mesa County, CO & -0.021$^{*}$ & -0.012$^{*}$ & -3.222$^{*}$ & -2.292$^{*}$ \\
 & & (-0.028, -0.017) & (-0.016, -0.009) & (-4.229, -2.534) & (-2.917, -1.772) \\

\midrule
17201 & Winnebago County, IL & -0.011$^{*}$ & -0.003$^{*}$ & -1.101$^{*}$ & -0.436$^{*}$ \\
 & & (-0.014, -0.009) & (-0.005, -0.001) & (-1.424, -0.705) & (-0.752, -0.195) \\
\midrule
21035 & Calloway County, KY & -0.014$^{*}$ & -0.026$^{*}$ & -1.711$^{*}$ & -2.352$^{*}$ \\
 & & (-0.029, -0.001) & (-0.042, -0.016) & (-3.573, -0.208) & (-3.878, -1.236) \\
\midrule
22057 & Lafourche Parish, LA & -0.035$^{*}$ & -0.050$^{*}$ & -0.665 & -2.821$^{*}$ \\
 & & (-0.051, -0.020) & (-0.066, -0.033) & (-2.513, 1.449) & (-4.598, -0.659) \\
\midrule
28121 & Rankin County, MS & -0.024$^{*}$ & -0.035$^{*}$ & -0.645 & -4.810$^{*}$ \\
 & & (-0.031, -0.017) & (-0.043, -0.029) & (-1.686, 0.355) & (-6.150, -3.601) \\
\midrule
34017 & Hudson County, NJ & -0.017$^{*}$ & -0.015$^{*}$ & -2.282$^{*}$ & -2.538$^{*}$ \\
 & & (-0.022, -0.012) & (-0.021, -0.011) & (-2.925, -1.686) & (-3.174, -1.953) \\
\midrule
36081 & Queens County, NY & 0.004$^{*}$ & 0.003$^{*}$ & -1.140$^{*}$ & -0.874$^{*}$ \\
 & & (0.001, 0.006) & (0.002, 0.005) & (-1.479, -0.756) & (-1.203, -0.591) \\
\midrule
39049 & Franklin County, OH & -0.016$^{*}$ & -0.016$^{*}$ & -11.023$^{*}$ & -4.352$^{*}$ \\
 & & (-0.017, -0.014) & (-0.017, -0.015) & (-11.329, -10.753) & (-4.544, -4.154) \\
\midrule
47021 & Cheatham County, TN & -0.005 & -0.029$^{*}$ & 0.340 & -2.647$^{*}$ \\
 & & (-0.015, 0.002) & (-0.044, -0.019) & (-0.827, 1.319) & (-4.804, -1.375) \\
\midrule
53029 & Island County, WA & -0.005$^{*}$ & -0.004$^{*}$ & -1.883$^{*}$ & -1.548$^{*}$ \\
 & & (-0.007, -0.003) & (-0.006, -0.002) & (-2.301, -1.488) & (-1.902, -1.176) \\
\bottomrule
\end{tabular}
\vspace{4pt}
\caption{Fairness (PRD) and Error (MAPE) Differences Between
Census and Baseline Models, Weighted and Unweighted.}
\label{tab:census_weighted_prd_mape}

\begin{tablenotes}[flushleft]
\scriptsize
\item \textit{Notes:} The table compares changes in assessment fairness and accuracy after the addition of Census characteristics for two model specifications: one where observations are unweighted, and one where observations are weighted by their sample inclusion probability during both model training and evaluation. ``Sample inclusion probability" is computed as the share of single-family homes in a block group that sell in a given year.
$^{*}$ denotes $p < 0.05$.
Confidence intervals are computed using a studentized bootstrap (95\%).
\end{tablenotes}

\end{table}

\end{document}